\documentclass[12pt]{article}
\usepackage{amsfonts}
\usepackage[left=2cm,right=2cm,top=2cm,bottom=2cm,bindingoffset=0cm]{geometry}
\usepackage{graphicx}
\usepackage{amsmath}
\usepackage{hyperref}
\usepackage{cite}
\usepackage{authblk}

\newcommand\beq{\begin{equation}}
\newcommand\eeq{\end{equation}}
\newcommand\nonum{\nonumber\\}

\title{\bf Comments on the adiabatic theorem}
\author[1, 2]{Dmitrii A. Trunin\thanks{\href{mailto:dmitriy.trunin@phystech.edu}{dmitriy.trunin@phystech.edu}}}
\affil[1]{Institutskii per, 9, Moscow Institute of Physics and Technology, 141700, Dolgoprudny, Russia}
\affil[2]{B. Cheremushkinskaya, 25, Institute for Theoretical and Experimental Physics, 117218, Moscow, Russia}
\date{\today}

\begin{document}

\maketitle

\begin{abstract}
We consider the simplest example of a nonstationary quantum system which is quantum mechanical oscillator with varying frequency and $\lambda \phi^4$ self-interaction. We calculate loop corrections to the Keldysh, retarded/advanced propagators and vertices using Schwinger--Keldysh diagrammatic technique and show that there is no physical secular growth of the loop corrections in the cases of constant and adiabatically varying frequency. This fact corresponds to the well-known adiabatic theorem in quantum mechanics. However, in the case of non-adiabatically varying frequency we obtain strong IR corrections to the Keldysh propagator which come from the `sunset' diagrams, grow with time indefinetely and indicate energy pumping into the system. It reveals itself via the change in time of the level population and of the anomalous quantum average.
\end{abstract}

\section{Introduction}

In high-energy particle physics one usually considers closed systems, i.e. systems parameters of which do not depend on external fields. Particularly, it is strongly believed that slowly varying external field cannot bring energy into system and change its parameters, so one can consider such a system as closed and use semi-classical approximation. However, sometimes this approximation breaks down. For example, many works on the Hawking radiation are based on semi-classical approximation~\cite{Hawking, Unruh}, but some recent papers show that non-stationarity of the background gravitational field leads to the secular growth of loop corrections to the correlation functions and suggest the breakdown of the perturbation theory~\cite{Akhmedov:2015}. The goal of this paper is to show exactly under what circumstances semi-classical (tree-level) approximation breaks down.

In non-stationary cases one has to use non-equilibrium Green's function approach which is also known as non-equilibrium diagrammatic technique or closed-time path or in-in formalism. This approach was initiated by Julian Schwinger~\cite{Schwinger} and further developed by L.~V.~Keldysh~\cite{Keldysh}. Schwinger--Keldysh approach has many applications in condensed matter physics~\cite{Garbrecht, Kamenev, Berges, Rammer, Calzetta, Millington, Mueller, Wang, Marino, Anastopoulos}, cosmology~\cite{Calzetta:1986, Polyakov:2012, Akhmedov:2012, Akhmedov:2013-phi3, Akhmedov:2013-phi4, Bascone:2017, Popov:2017, Glavan, Petri, Giddings, Prokopec:2003}, ultra relativistic heavy ion collisions~\cite{Prokopec:2004, Dusling, Epelbaum, Leonidov:2014, Leonidov:2016}, non-stationary phenomena in the strong background field~\cite{Nikishov, Narozhnyi, Akhmedov:2014-1, Akhmedov:2014-2} and so on.

Here we consider the simplest example of such a system which is non-linear quantum mechanical oscillator with varying frequency and $\lambda \phi^4$ interaction. Since many systems (e.g. phononic systems at the ballistic level) can be thought of as coupled oscillators, and an infinite number of harmonic oscillators make up a free quantum field, this problem is fundamental to the non-equilibrium Green's function approach. On the one hand, we show here that in non-adiabatically varying frequency case level population and anomalous quantum average of the system receive large corrections, which emphasizes that adiabaticity of the change in parameters plays an important role even in such a simple system. On the other hand, we also show that the case of adiabatically varying frequency is literally indistinguishable from the case of constant frequency, i.e. stationary case. This fact corresponds to the well-known adiabatic theorem in quantum mechanics~\cite{Born, Kato}.

It is worth stressing here that in quantum mechanics one usually considers measurements, i.e. opens up the system by an external device and measures the collapsed state. In other words, one considers the wave function of the system and calculates the probabilities of transition between its eigen-states~\cite{Zel'dovich, Popov:1, Popov:2}. However, here we restrict ourselves to a quasi-closed quantum mechanical system, i.e. we study the evolution of the self-interacting quantum field in the absence of interactions with the outer world apart from those indirect interactions which change the frequency. I.e. here we consider quantum mechanics as a simplest example for quantum field theory. Note, however, that in low dimensions those IR phenomena that we consider in this paper are frequently stronger than in higher dimensions.

We develop a diagrammatic technique to calculate the loop corrections to the so-called Keldysh, retarded and advanced propagators similarly to the quantum field theory case and check the results with direct calculations using the decomposition of the evolution operator. Surprisingly, at first glance, corrections to the Keldysh propagator do grow with time even in the cases of adiabatically varying and constant frequency. At the same time, this propagator is related to the level population and anomalous quantum average~\cite{Garbrecht}, so one should expect that it will not receive large corrections in these cases due to the adiabatic theorem. Indeed, we show that the growth can be removed by a modification of the vacuum state which restores the vacuum state of the free Hamiltonian on the past and future infinities. We stress that there is no such a growth in quantum field theory, i.e. it is peculiar to quantum mechanics. 

Nevertheless, in the case of non-adiabatically varying frequency the growth is more fundamental. Despite the fact that one can hide `tadpole' diagrams into a renormalization of the frequency and remove the growth in the first order in $\lambda$, `sunset' diagrams do grow with time anyway. Moreover, they behave as $\sim (\lambda T)^2$ as the average time of the two-point function $T \rightarrow \infty$, i.e. the IR divergence is stronger than in quantum field theory of higher dimensions which is $\sim \lambda^2 T$. This is a common property of QFTs of low dimensions~\cite{Alexeev, Astrahantsev}. Finally, we stess that one should have expected such a growth in this case because the non-adiabaticity of the time dependence of the frequency brings energy into the system.

It is worth stressing that there are differences between quantum mechanics and quantum field theory. Namely, usually one uses Wick's theorem to build a diagrammatic technique and calculate corrections to the correlation functions in an arbitrary state~\cite{Matsubara, Landau:vol9, Landau:vol10} and then deduces the kinetic equation using this technique~\cite{Kamenev, Berges, Rammer, Akhmedov:2013-phi3, Akhmedov:2013-phi4, Polyakov:2012, Akhmedov:2012}. At the same time, in quantum mechanics Wick's theorem works only for the vacuum and thermal states due to the absence of the spatial volume~\cite{Evans, Wick, Peskin, Fetter}. This means that we cannot sum up the leading contribution in all loops, derive the kinetic equation and find the correct time evolution of the level population using the standard techniques.

This paper is organized as follows. In Section~\ref{sec:constfreq} we consider the oscillator with constant frequency and adiabatically turning $\lambda \phi^4$ interaction, discuss the physical origin of the secular growth and differences from quantum field theory. In Section~\ref{sec:varfreq} we turn to the non-stationary case and consider adiabatically and non-adiabatically varying frequency cases separately. In Section~\ref{sec:diagrams} we introduce the diagrammatic technique, show that it gives correct results in the first lower loop-corrections, calculate corrections to the vertices and discuss a renormalization of the frequency. Finally, we conclude in Section~\ref{sec:conclusion}.

\section{Constant frequency case}
\label{sec:constfreq}

\subsection{Setup of the problem}

To set up the notations we start our considerations with the nonlinear $\lambda \phi^4$ oscillator with constant frequency~\cite{Kamenev, Wang}:
\beq
\label{eq:QM-action}
S = \int dt \Big[ \frac{1}{2} \dot \phi^2(t) - \frac{\omega^2}{2} \phi^2(t) - \frac{\lambda}{4} \phi^4(t) \Big].
\eeq
The Lagrangian and the Hamiltonian of the theory are:
\beq L = \frac{1}{2} \dot \phi^2 - \frac{\omega^2}{2} \phi^2  - \frac{\lambda}{4} \phi^4 \quad {\rm and} \quad H = \frac{1}{2} \dot \phi^2 + \frac{\omega^2}{2} \phi^2 + \frac{\lambda}{4} \phi^4. \eeq
Introducing ladder operators $a$ and $a^+$ and representing the real field $\phi(t)$ as
\beq \phi(t) = \frac{1}{\sqrt{2\omega}} \Big[ a e^{i \omega t} + a^+ e^{-i \omega t} \Big] = a f(t) + a^+ f^*(t), \eeq
we obtain that the free Hamiltonian has the standard form:
\beq H_0(t) = \frac{1}{2} \dot \phi^2 + \frac{\omega^2}{2} \phi^2 = \omega \Big( a^+ a + \frac{1}{2} \Big). \eeq
Here for short we have introduced the following notation:
\beq
\label{eq:simplemode}
f(t) = \frac{1}{\sqrt{2\omega}} e^{i \omega t}.
\eeq
Hence, the free theory evolution operator is:
\beq U_0(t_2, t_1) = T\exp\Big[ -i \int_{t_1}^{t_2} d\tau H_0(\tau) \Big] = \exp \Big[ -i (t_2 - t_1) H_0 \Big]. \eeq
Self-interaction operator in the interaction picture:
\begin{align}
\label{eq:interaction}
V(t) &= U_0^+(t, t_0) \Big(\frac{\lambda}{4} \phi^4 \Big) U_0(t, t_0) = \frac{\lambda}{4} \Big( U_0^+(t, t_0) \phi U_0(t, t_0) \Big)^4 = \frac{\lambda}{4} \phi^4(t) = \nonum &= \frac{\lambda}{4} \Big( a^4 f^4(t) + (a^+)^4 f^{*4}(t) + 2 B f^2(t) |f(t)|^2 + 2 B^+ f^{*2}(t) |f(t)|^2 + 3 A |f(t)|^4 \Big).
\end{align}
Here $t_0$ is the time after which the self-interaction $\lambda \phi^4$ is adiabatically turned on. It is worth stressing here that in the case of constant frequency all the calculated quantities depend on the difference $t - t_0$ instead of $t$ and $t_0$ separately due to the translational invariance of the action. Hence, we set $t_0 = 0$ in this section without loss of generality. Also we introduce for convenience the following notations:
\beq
\label{eq:AB}
A = a a^+ a a^+ + a^+ a a^+ a, \quad B = a^3 a^+ + a^+ a^3.
\eeq
These operators have the following properties:
\begin{gather}
A | n \rangle = (2n^2 + 2n + 1) | n \rangle, \quad A^+ | n \rangle = A | n \rangle = (2n^2 + 2n + 1) | n \rangle, \nonum B | n \rangle = (2n - 1) \sqrt{n(n - 1)} | n - 2 \rangle, \quad B^+ | n \rangle = (2n + 3) \sqrt{(n + 1)(n + 2)} | n + 2 \rangle.
\end{gather}
Finally, we introduce the notations (we assume that $t_1 \ge t_2$):
\beq
\label{eq:times}
T = \frac{1}{2} (t_1 + t_2), \quad \tau = t_1 - t_2.
\eeq
Thus, evolution operator in the interaction picture is as follows:
\begin{align}
\label{eq:evolution}
U(t_b, t_a) &= T\exp\Big[ -i \int_{t_a}^{t_b} d\eta V(\eta) \Big] = 1 - i \int_{t_a}^{t_b} d\eta V(\eta) + (-i)^2 \int_{t_a}^{t_b} d\eta V(\eta) \int_{t_a}^{\eta} d\xi V(\xi) + \cdots \equiv \nonum &\equiv 1 + U_1(t_b, t_a) + U_2(t_b, t_a) + \cdots.
\end{align}
Obviously, it has the following properties:
\beq U(t_a, t_b) U(t_b, t_c) = U(t_a, t_c), \quad U^+(t_a, t_b) = U(t_b, t_a). \eeq
It is easy to calculate the first order correction for the evolution operator explicitly:
\begin{align}
U_1(t_b, t_a) = \frac{\lambda}{64\omega^3} &\Big[ a^{+4} \Big( e^{-4 i \omega t_b} - e^{-4 i \omega t_a}\Big) - a^4 \Big( e^{4 i \omega t_b} - e^{4 i \omega t_a} \Big) + \nonum &+ 4 B^+ \Big( e^{-2 i \omega t_b} - e^{-2 i \omega t_a} \Big) - 4 B \Big( e^{2 i \omega t_b} - e^{2 i \omega t_a} \Big) - 12 A i \omega (t_b - t_a) \Big].
\end{align}
Unfortunately, the number of terms contained in the corrections to the evolution operator grows rapidly with the increase of the order of $\lambda$. For example, the second order contains more than one hundred different terms. However, most of them may be omitted if we look for the leading growing in time expressions. Throughout this paper we consider the limit $T \rightarrow \infty$, $\tau = \text{const}$:
\begin{align}
U_1(t_{1/2}, 0) &= \Big( \frac{-3 i \lambda}{16\omega^3} \Big) \cdot \omega T \cdot A + O(T^0), \nonum
U_1(t_2, t_1) &= O(T^0), \nonum
U_2(t_{1/2}, 0) &= \frac{1}{2} \Big( \frac{-3 i \lambda}{16\omega^3} \Big)^2 \cdot \omega^2 (T^2 \pm T \tau) \cdot A^2 - \nonum &- \frac{i}{12} \Big( \frac{-3 i \lambda}{16\omega^3} \Big)^2 \cdot \omega T \cdot \Big[ \frac{1}{3} \Big( a^{+4} a^4 - a^4 a^{+4} \Big) + \frac{8}{3} \Big( B^+ B - B B^+ \Big) + \nonum &+ \Big( a^4 A e^{4 i \omega t_{1/2}} - A a^4 \Big) + \Big( 4 B A e^{2 i \omega t_{1/2}} - 4 A B \Big) - \nonum &- \Big( a^{+4} A e^{-4 i \omega t_{1/2}} - A a^{+4} \Big) - \Big( 4 B^+ A e^{-2 i \omega t_{1/2}} - 4 A B^+ \Big) \Big] + O(T^0), \nonum
U_2(t_2, t_1) &= O(T^0) \cdot O(\tau).
\end{align}
This limit corresponds to the case that both points of the two-point correlation function $\langle \phi(t_1) \phi(t_2) \rangle$ are taken to the future infinity while the time difference between them is kept fixed. We consider such a limit to check the evolution of the state due to the self-interaction.

\subsection{Two-point functions and perturbative corrections}
\label{subsec:constfreq}

Now let us calculate the correlation (Wightman) function of two fields in the $n$-th eigen-state of the free Hamiltonian $H_0|n\rangle = \Big( n + \frac{1}{2} \Big) | n \rangle$:
\begin{align}
\label{eq:correlator}
C(1, 2) &= \langle \phi(t_1) \phi(t_2) \rangle = \langle n| U^+(t_1, 0) \phi(t_1) U^+(t, t_1) U(t, t_2) \phi(t_2) U(t_2, 0) |n \rangle = \nonum & = \Big\langle n \Big| \Big[ 1 + U_1(0, t_1) + U_2(0, t_1) + \cdots \Big] \phi(t_1) \Big[ 1 + U_1(t_1, t_2) + U_2(t_1, t_2) + \cdots \Big] \phi(t_2) \cdot \nonum &\phantom{=}\cdot \Big[ 1 + U_1(t_2, 0) + U_2(t_2, 0) + \cdots \Big] \Big| n \Big\rangle = C_0(t_1, t_2) + C_1(t_1, t_2) + C_2(t_1, t_2) + \cdots,
\end{align}
where we define (recall that we set $t_0 = 0$):
\begin{align}
\label{eq:correlators}
C_0(1, 2) &\equiv \langle n | \phi_1 \phi_2 | n \rangle, \nonum
C_1(1, 2) &\equiv \langle n | U_1(0, 1) \phi_1 \phi_2 + \phi_1 U_1(1, 2) \phi_2 + \phi_1 \phi_2 U_1(2, 0) | n \rangle, \nonum
C_2(1, 2) &\equiv C_2^1(1, 2) + C_2^2(1, 2), \nonum
C_2^1(1, 2) &\equiv \langle n | U_1(0, 1) \phi_1 U_1(1, 2) \phi_2 + U_1(0, 1) \phi_1 \phi_2 U_1(2, 0) + \phi_1 U_1(1, 2) \phi_2 U_1(2, 0) | n \rangle, \nonum
C_2^2(1, 2) &\equiv \langle n | U_2(0, 1) \phi_1 \phi_2 + \phi_1 U_2(1, 2) \phi_2 + \phi_1 \phi_2 U_2(2, 0) | n \rangle.
\end{align}
Here we denoted $\phi(t_a) \equiv \phi_a$ and $C(t_a, t_b) \equiv C(a, b)$ for short. It is not difficult to calculate the correlators $C_0(1, 2)$ and $C_1(1, 2)$. At the same time, the difficulty of the calculation of $C_2(1, 2)$ increases significantly while most of the terms here are suppressed in the limit $T \rightarrow \infty$, $\tau = \text{const}$. We will keep only the leading terms in such a limit.

Thus, at the zeroth order:
\beq
\label{eq:C0}
C_0(1, 2) = n f_1^* f_2 + (n + 1) f_1 f_2^* = \frac{1}{2\omega} \Big( n \exp(-i \omega \tau) + (n + 1)\exp(i \omega \tau)\Big).
\eeq
The leading expressions of the first order cancel each other due to the properties of the operator $A$ and as a result:
\begin{align}
\label{eq:C1}
C_1(1, 2) &= \frac{3 i \lambda}{16\omega^3} \cdot \omega T \cdot \langle A \phi_1 \phi_2 \rangle - \frac{3 i \lambda}{16\omega^3} \cdot \omega T \cdot \langle \phi_1 \phi_2 A \rangle + O(T^0) = O(T^0),
\end{align}
i.e. $C_1$ does not grow as $T \rightarrow \infty$.

The second order correction in an arbitrary state contain the huge number of terms. Therefore, we first calculate it in the vacuum state of the free Hamiltonian: $H_0 | 0 \rangle = \frac{1}{2} | 0 \rangle$, and then generalize the result. The first part is as follows:
\begin{align}
C_2^1(1, 2) &= \Big( \frac{3 \lambda}{16\omega^3} \Big)^2 \cdot (\omega T)^2 \cdot \langle A \phi_1 \phi_2 A \rangle - \nonum &- \Big( \frac{3 \lambda}{16\omega^3} \Big)^2 \cdot \frac{i \omega T}{3} \cdot \Big[ \langle A a^2 B^+ e^{2 i \omega t_2} - A a^2 B^+ \rangle f_1 f_2 - \langle A B a^{+2} e^{-2 i \omega t_1} - A B a^{+2} \rangle f_1^* f_2^* \Big] + \nonum &+ O(T^0),
\end{align}
the second part:
\begin{align}
C_2^2(1, 2) &= -\frac{1}{2} \Big( \frac{3 \lambda}{16\omega^2} \Big)^2 \cdot \Big[ (T^2 + T \tau) \langle A^2 \phi_1 \phi_2 \rangle + (T^2 - T \tau) \langle \phi_1 \phi_2 A^2 \rangle \Big] + \nonum &+ \Big( \frac{3 \lambda}{16\omega^3} \Big)^2 \cdot \frac{i \omega T}{3} \cdot \Big[ \langle a^2 B^+ A e^{2 i \omega t_2} - a^2 A B^+ \rangle f_1 f_2 - \langle A B a^{+2} e^{-2 i \omega t_1} - B A a^{+2} \rangle f_1^* f_2^* \Big] + \nonum &+ O(T^0).
\end{align}
The leading orders of these expressions $\sim T^2$ cancel each other. However, $C_2$ grows linearly as $T \rightarrow \infty$:
\begin{align}
\label{eq:C2-const}
C_2(1, 2) &= -\frac{i \omega T}{3} \Big( \frac{3 \lambda}{16\omega^3} \Big)^2 \langle a^2 A B^+ - A a^2 B^+ \rangle \Big( f_1 f_2 - f_1^* f_2^* \Big) + O(T^0) = \nonum &= 27 i \cdot \frac{\lambda^2 T}{(2 \omega)^5} \Big( f_1^* f_2^* - f_1 f_2 \Big) + O(T^0).
\end{align}
Performing similar calculation one can find that in an arbitrary $n$-eigenstate of the free Hamiltonian:
\beq C_2(1, 2) = (48 n^3 + 72 n^2 + 78 n + 27) \cdot \frac{i \lambda^2 T}{(2 \omega)^5} \Big( f_1^* f_2^* - f_1 f_2 \Big) + O(T^0). \eeq
To calculate $\langle \phi(t_2)\phi(t_1) \rangle$ we should simply change $t_1 \leftrightarrow t_2$. It corresponds to $T \rightarrow T$ and $\tau \rightarrow -\tau$. Then we may find the expressions for the Keldysh $D^K(t_1, t_2)$ and retarded/advanced (R/A) $D^{R/A}(t_1, t_2)$ propagators:
\begin{align}
\label{eq:propagators}
D^K(t_1, t_2) &= \frac{1}{2} \big\langle \lbrace \phi(t_1), \phi(t_2) \rbrace \big\rangle, \nonum
D^{R/A}(t_1, t_2) &= \pm \theta(\pm t_1 \mp t_2) \big\langle [ \phi(t_1), \phi(t_2) ] \big\rangle.
\end{align}
We notice here that
\beq D^A(t_1, t_2) = D^R(t_2, t_1), \quad \text{i.e.} \quad D^A(T, \tau) = D^R(T, -\tau). \eeq
This means that advanced and retarded propagators behave identically in the limit in question and we can calculate only the retarded one. So we immediately obtain that in the vacuum state:
\begin{align}
\label{eq:prop-const}
D_0^K(T, \tau) &= \frac{1}{2} \big( f_1 f_2^* + f_1^* f_2 \big) = \frac{1}{2\omega} \cos(\omega \tau), \nonum
D_1^K(T, \tau) &= O(T^0), \nonum
D_2^K(T, \tau) &= 27 i \cdot \frac{\lambda^2 T}{(2 \omega)^5} \Big( f_1^* f_2^* - f_1 f_2 \Big) = 54 \frac{\lambda^2 T}{(2 \omega)^6} \sin(2 \omega T) + O(T^0), \nonum
D_0^R(T, \tau) &= \theta(\tau) \big( f_1 f_2^* - f_1^* f_2 \big) = \theta(\tau)\frac{i}{\omega} \sin(\omega \tau), \nonum
D_1^R(T, \tau) &= D_2^R(T, \tau) = O(T^0).
\end{align}
To conclude, we have obtained that corrections to the R/A propagators do not grow in the limit $T \rightarrow \infty$, $\tau = \text{const}$, as well as the first order correction to the Keldysh propagator. At the same time, the second order correction to the Keldysh propagator does grow in this limit. In other words, we have obtained the secular growth in the second order loop correction.

Moreover, one can associate the Keldysh propagator with energy level population $n = \langle a^+ a \rangle$ and anomalous quantum average $\kappa = \langle a a \rangle$, directly calculating the expression~\eqref{eq:propagators}:
\beq
\label{eq:nk}
D^K(t_1, t_2) = \Big( n(T) + \frac{1}{2} \Big) f_1 f_2^* + \kappa(T) f_1 f_2 + h.c.
\eeq
Hence, the secular growth of the Keldysh propagator is connected with the growth of such an expressions~\cite{Garbrecht}. However, in the following subsection~\ref{subsec:causes} we will show that this growth can be removed by a redefinition of the vacuum state.

\subsection{The physical origin of the secular growth}
\label{subsec:causes}

One should expect that neither corrections to the R/A nor those to the Keldysh propagator grow with the time $T$ because of the time independence of $\omega$ and adiabaticity of the time dependence of $\lambda$. In fact, neither level population nor anomalous quantum average which are components of the Keldysh propagator should receive large corrections in this case due to the adiabatic theorem~\cite{Born, Kato}. In this subsection we will clarify the roots of the found growth which seems puzzling in the light of the theorem. To do this let us modify the vacuum state of the free Hamiltonian $H_0$:
\beq | \Omega \rangle = \mu_0 | 0 \rangle + \mu_1 | 1 \rangle + \mu_2 | 2 \rangle + \cdots. \eeq 
The normalization relation of the state $\langle \Omega | \Omega \rangle = 1$ reads:
\beq |\mu_0|^2 + |\mu_1|^2 + |\mu_2|^2 + \cdots = 1, \eeq
and the correlation function changes as follows:
\begin{align}
\label{eq:Comega}
C^\Omega(1, 2) &= \langle \Omega | U^+(1, 0) \phi_1 U(1, 2) \phi_2 U(2, 0) | \Omega \rangle = \langle \Omega | \cdots | \Omega \rangle = \nonum &= |\mu_0|^2 \langle 0 | \cdots | 0 \rangle + |\mu_1|^2 \langle 1 | \cdots | 1 \rangle + |\mu_2|^2 \langle 2 | \cdots | 2 \rangle + \mu_0 \mu_2^* \langle 2 | \cdots | 0 \rangle + \mu_0^* \mu_2 \langle 0 | \cdots | 2 \rangle + \cdots.
\end{align}
Here we restrict ourselves to the first two terms in the expansion of the state because this is enough to understand the behavior in the $\lambda^2$ order. We call the first three terms in~\eqref{eq:Comega} as `diagonal' and the last two terms as `off-diagonal'. Also we expand the correlation function in powers of $\lambda$ in the way similar to~\eqref{eq:correlators}: $C^\Omega(1, 2) = C_0^\Omega(1, 2) + C_1^\Omega(1, 2) + C_2^\Omega(1, 2) + \cdots$, replacing the state~$| n \rangle$ with the state~$| \Omega \rangle$. We remind that diagonal contributions to the $C_0$ and $C_1$ corrections have been calculated in the previous subsection~\eqref{eq:C0},~\eqref{eq:C1}. Substituting these expressions into the formula~\eqref{eq:Comega} we obtain:
\begin{align}
C_0^\Omega(1, 2) &= |\mu_0|^2 f_1 f_2^* + |\mu_1|^2 \big( 2 f_1 f_2^* + f_1^* f_2 \big) + \nonum &+ |\mu_2|^2 \big( 3 f_1 f_2^* + 2 f_1^* f_2 \big) + \sqrt{2} \big( \mu_0 \mu_2^* f_1^* f_2^* + \mu_0^* \mu_2 f_1 f_2 \big) + \cdots, \nonum
C_1^\Omega(1, 2) &= 9 \sqrt{2} \frac{i \lambda T}{(2 \omega)^2} \cdot \big( \mu_0 \mu_2^* f_1^* f_2 ^* - \mu_0^* \mu_2 f_1 f_2 \big) + O(T^0) + \cdots.
\end{align}
Note that correction $C_1^\Omega$ is proportional to $T$ due to the off-diagonal elements from~\eqref{eq:Comega}. The second-order diagonal correction $C_2^\Omega$ equals to
\beq C_2^{\Omega, \text{diagonal}}(1, 2) = \big( 27 |\mu_0|^2 + 225 |\mu_1|^2 + 855 |\mu_2|^2 + \cdots \big) \cdot \frac{\lambda^2 T}{(2 \omega)^5} \cdot \big( f_1^* f_2^* - f_1 f_2 \big) + O(T^0), \eeq
and off-diagonal contribution is as follows:
\beq C_2^{\Omega, \text{off-diagonal}}(1, 2) = - \frac{81}{\sqrt{2}} \frac{\lambda^2 T^2}{(2 \omega)^4} \cdot \big( \mu_0 \mu_2^* f_1^* f_2 ^* + \mu_0^* \mu_2 f_1 f_2 \big) + O\big(\mu_0 \mu_2 T\big) + \cdots. \eeq
If we choose the coefficients $\mu_i$ in the following way:
\beq
\label{eq:mus}
\mu_0 = 1 - \frac{1}{(6 \omega T)^2}, \quad \mu_1 = 0, \quad \mu_2 = \frac{-i}{3 \sqrt{2} \cdot \omega T},
\eeq
we will obtain that both the first and the second corrections to the correlation function $C^\Omega$ do not depend on time $T$ in the leading expressions as $T \rightarrow \infty$. At the same time, one can see that $\mu_0 \rightarrow 1$ and $\mu_1, \mu_2 \rightarrow 0$ in the limit $T \rightarrow \infty$, i.e. the new vacuum state tends to the vacuum state of the free Hamiltonian $| \Omega \rangle \rightarrow | 0 \rangle$. Thus, one can remove the secular growth~\eqref{eq:C2-const} by a redefinition of the vacuum state on the past and future infinities. This is due to the rapid growth of the off-diagonal expressions which cancel the diagonal contribution to $C^\Omega$ in the same order in $\lambda$. This is why coefficients~\eqref{eq:mus} do not depend on $\lambda$. In general, one can check that in the $\lambda^n$ order ($n \ge 1$) the diagonal contribution is $C_n^{\Omega, \text{diagonal}} \sim T^{n-1}$ and off-diagonal contribution is $C_n^{\Omega, \text{off-diagonal}} \sim T^n$. Hence, one can make the similar change and remove the secular growth in an each order. Finally, let us stress that the expressions~\eqref{eq:mus} is valid only in the limit $T \rightarrow \infty$, and there is no divergence at $T \rightarrow 0$.

Thus, the growth of the second order correction~\eqref{eq:C2-const} is connected with the off-diagonal elements coming from the evolved vacuum state of the free theory:
\begin{align}
U(t, 0) | 0 \rangle &= | 0 \rangle - \frac{3 \lambda t}{16 \omega^2} | 0 \rangle - \frac{1}{2} \Big( \frac{3 \lambda t}{16 \omega^2} \Big)^2 | 0 \rangle - \nonum &- \frac{i}{12} \Big( \frac{3 \lambda}{16 \omega^3} \Big)^2 \omega t \Big[ 56 | 0 \rangle + 12 \sqrt{2} \Big( e^{-2 i \omega t} - 13 \Big) | 2 \rangle + \sqrt{24} \Big( e^{-4 i \omega t} - 41 \Big) | 4 \rangle \Big].
\end{align}
Indeed, only the expectation values that contain the operator $B$ survive in the expression~\eqref{eq:C2-const} which indicates the evolution of the vacuum state $| 0 \rangle$ into the state $| 2 \rangle$. In other words, we have the growth in the second-order loop correction to the Keldysh propagator because of `hopping' out of the vacuum state of the free Hamiltonian due to the presence of the self-interaction $\lambda \phi^4$.

Let us stress that similar higher order corrections to the correlation functions also appear in quantum field theory. However, usually one removes this growth by a renormalization of the mass and of the coupling constant of the theory instead of the redefinition of the ground vacuum state~\cite{Peskin, Fetter, Wilson}. We will show in the subsection~\ref{subsec:renorm} that one can make such renormalization to remove some growing corrections in the case of non-adiabatically varying frequency, but this approach does not work in constant~\eqref{eq:QM-action} and adiabatically varying~\eqref{eq:varaction} frequency cases. Indeed, second order correction to the correlation function~\eqref{eq:C2-const} contains terms of the form $f_1 f_2$ and $f_1^* f_2^*$ which are absent in the zeroth order~\eqref{eq:C0}. Hence, one cannot obtain such terms by a change of the frequency in~\eqref{eq:C0} and a redefinition of the vacuum state is required.

Finally, let us also stress that such contributions as~\eqref{eq:C2-const} are forbidden in quantum field theory by the energy-momentum conservation law~\cite{Landau:vol2}.

\subsection{Wick's theorem}
\label{subsec:Wick}

In this subsection we consider the real massive scalar field theory in $3+1$ dimensions and discuss the differences from the quantum mechanics in $0+1$ dimensions~\eqref{eq:QM-action}. Here we just repeat the standard textbook reasoning to stress the difference of the quantum mechanics from the quantum field theory. Namely, we consider the following action:
\beq
\label{eq:QFT}
S = \int d^4 x \Big[ \frac{1}{2} (\partial_\mu \phi)^2 - \frac{m^2}{2} \phi^2 - \frac{\lambda}{4} \phi^4 \Big].
\eeq
As usual, we introduce the interaction picture: $\phi(x) = e^{i H_0 t} \phi(\vec{x}) e^{-i H_0 t}$, where $H_0$ is the free Hamiltonian and $U_0(t, 0) = e^{-i H_0 t}$ is the free theory evolution operator. Then we quantize the field at a given moment of time $t_0$ as follows~\cite{Peskin}:
\beq \phi(t_0, \vec{x}) = \int \frac{d^3 p}{(2\pi)^3} \frac{1}{\sqrt{2 E_p}} \Big( a_p e^{i \vec{p} \vec{x}} + a_p^+ e^{-i \vec{p} \vec{x}} \Big), \eeq
and for an arbitrary time we obtain:
\beq
\label{eq:QFT-decomposition}
\phi(t, \vec{x}) = e^{i H_0 (t - t_0)} \phi(t_0, \vec{x}) e^{-i H_0 (t - t_0)} = \int \frac{d^3 p}{(2\pi)^3} \frac{1}{\sqrt{2 E_p}} \Big( a_p e^{-i p x} + a_p^+ e^{i p x} \Big).
\eeq
Here the creation $a_p^+$ and annihilation $a_p$ operators obey the following commutation relations:
\beq
\label{eq:commutators}
\big[ a_p, a_q^+ \big] = \delta^{(3)}(\vec{p} - \vec{q}), \quad \big[ a_p, a_q \big] = \big[ a_p^+, a_q^+ \big] = 0.
\eeq
Mathematically, it is more convenient first to consider the field in a finite volume $\mathcal{V}$ and then take the limit $\mathcal{V} \rightarrow \infty$:
\beq
\label{eq:pos+neg}
\phi(x) = \phi^+(x) + \phi^-(x) = \sum_{p} \frac{1}{\sqrt{2 E_p \mathcal{V}}} a_p e^{-i p x} + \sum_{p} \frac{1}{\sqrt{2 E_p \mathcal{V}}} a_p^+ e^{i p x}.
\eeq
The main property of the quantum field theory is the Wick's theorem which allows one to calculate corrections to correlation functions~\cite{Matsubara, Landau:vol9, Landau:vol10, Evans}. However, one should cautiously use this theorem in quantum mechanics since its proof involves the averaging over the spatial volume. Indeed, let us for example consider the following average of the four fields:
\beq
\label{eq:Wick}
\langle \phi^+(x_1) \phi^+(x_2) \phi^-(x_3) \phi^-(x_4) \rangle = \frac{1}{\mathcal{V}^2} \sum_{p_1 \cdots p_4} \langle a_{p_1} a_{p_2} a_{p_3}^+ a_{p_4}^+ \rangle \exp(\cdots),
\eeq
where $\exp(\cdots)$ denotes the product of the corresponding exponential functions and summation is performed over all the four momenta. In this sum only such terms are non-zero which contain the equal number of operators $a_p$ and $a_p^+$. So one can reduce the sum over all the four momenta to the sum of two independent momenta and then split the averages:
\beq \langle a_{p_1} a_{p_2} a_{p_3}^+ a_{p_4}^+ \rangle \rightarrow \langle a_{p_1} a_{p_3}^+ \rangle \langle a_{p_2} a_{p_4}^+ \rangle + \langle a_{p_1} a_{p_4}^+ \rangle \langle a_{p_2} a_{p_3}^+ \rangle \rightarrow ( \delta_{p_1, p_3} \delta_{p_2, p_4} + \delta_{p_1, p_4} \delta_{p_2, p_3} ) \langle a_{p_1} a_{p_1}^+ \rangle \langle a_{p_2} a_{p_2}^+ \rangle. \eeq
In this case we obtain that the expression remains finite in the limit $\mathcal{V} \rightarrow \infty$ since both sums are replaced with the integrations $\sum_{p} \rightarrow \int \frac{\mathcal{V} d^3 p}{(2 \pi)^3}$:
\beq
\label{eq:Wick-non-zero}
\frac{1}{\mathcal{V}^2} \sum_{p_1, p_2} \langle a_{p_1} a_{p_1}^+ \rangle \langle a_{p_2} a_{p_2}^+ \rangle \exp(\cdots) \longrightarrow \int \frac{d^3 p_1 d^3 p_2}{(2 \pi)^6} \langle a_{p_1} a_{p_1}^+ \rangle \langle a_{p_2} a_{p_2}^+ \rangle \exp(\cdots).
\eeq
On the other hand, there are other non-zero terms in the sum~\eqref{eq:Wick} which are obtained when all the four momenta are equal. However, they tend to zero in the limit $\mathcal{V} \rightarrow \infty$: 
\beq
\label{eq:Wick-zero}
\frac{1}{\mathcal{V}^2} \sum_{p} \langle a_p a_p a_p^+ a_p^+ \rangle \exp(\cdots) \longrightarrow \frac{1}{\mathcal{V}} \int \frac{d^3 p}{(2 \pi)^3} \langle a_p a_p a_p^+ a_p^+ \rangle \exp(\cdots) \sim \frac{1}{\mathcal{V}} \rightarrow 0,
\eeq
and hence we can reduce the sum~\eqref{eq:Wick} to the sum of the products of the pair contractions. Thus, the $4$-point correlation function decomposes as follows:
\beq
\label{eq:Wick-4-point}
\langle \phi_1 \phi_2 \phi_3 \phi_4 \rangle = \langle \phi_1 \phi_2 \rangle \langle \phi_3 \phi_4 \rangle + \langle \phi_1 \phi_3 \rangle \langle \phi_2 \phi_4 \rangle + \langle \phi_1 \phi_4 \rangle \langle \phi_2 \phi_3 \rangle,
\eeq
where we denoted $\phi(x_a) \equiv \phi_a$ for short. One can apply similar reasoning to an arbitrary $n$-point correlation functions.

Let us emphasize that we do not actually use the fact that the averaging is performed over the vacuum state. In fact, one can consider averages over arbitrary coherent states as well. In the latter case the presence of the infinite volume also allows one to eliminate terms of the form~\eqref{eq:Wick-zero}. Hence, only the pair averages do survive.

Also let us stress here that one does not need to set $\mathcal{V} \rightarrow \infty$ to prove Wick's theorem, i.e. any $n$-point Green function in the quantum field theory can be decomposed exactly into the product of two-point Green functions even in the case $\mathcal{V} < \infty$. E.g. one can find the proof in the paper~\cite{Evans}. However, in this paper only vacuum or thermal expectation values are considered whereas the proof we reproduce here works for averages over arbitrary states as well.

Moreover, one can prove Wick's theorem avoiding spatial volumes at all~\cite{Wick, Peskin, Fetter}. However, in this case one should restrict oneself to the vacuum averages only. Indeed, let us check by induction the following statement:
\beq
\label{eq:Wick-vacuum}
\phi_1 \phi_2 \cdots \phi_m = \mathcal{N}\Big\{ \phi_1 \phi_2 \cdots \phi_m + \Big( \begin{smallmatrix} \text{all possible} \\ \text{pair contractions} \end{smallmatrix} \Big) \Big\},
\eeq
where $\mathcal{N}\{\cdots\}$ is normal ordering. We define the contraction of the fields $\phi_a^\bullet \phi_b^\bullet \equiv [ \phi_a^+, \phi_b^- ]$. The contraction defined in this way is a c-number, and in the vacuum state it is equal to the average $\phi_a^\bullet \phi_b^\bullet = \langle \phi_a \phi_b \rangle$ (this is not the case in an arbitrary state, however). In the case of two fields equality~\eqref{eq:Wick-vacuum} is obviously satisfied. Then we need the following identity:
\begin{align}
\underbrace{\phi_1^- \phi_2^- \cdots \phi_{i-1}^- \phi_i^+ \phi_{i+1}^+ \cdots \phi_m^+}_{\text{normal-ordered product}} \phi_{m+1}^- &= \phi_1^- \phi_2^- \cdots \phi_{i-1}^- [\phi_i^+, \phi_{m+1}^-] \phi_{i+1}^+ \cdots \phi_m^+ + \nonum &+ \phi_1^- \phi_2^- \cdots \phi_{i-1}^- \phi_i^+  [\phi_{i+1}^+, \phi_{m+1}^-] \cdots \phi_m^+ + \nonum &+ \cdots + \nonum &+ \phi_1^- \phi_2^- \cdots \phi_{i-1}^- \phi_i^+ \phi_{i+1}^+ \cdots [ \phi_m^+, \phi_{m+1}^- ].
\end{align}
Hence, multiplying both sides of~\eqref{eq:Wick-vacuum} by the operator $\phi_{m+1}$, splitting $\phi_{m+1}$ on positive and negative parts~\eqref{eq:pos+neg} and rearranging operators on the r.h.s. to obtain normal-ordered products, one makes an induction step. Finally, when one averages the identity~\eqref{eq:Wick-vacuum} over the vacuum state, only fully contracted expressions do survive due to the property of the normal ordering. So one obtains the decomposition of the form~\eqref{eq:Wick-4-point}.

This means that one can use Wick's theorem in quantum mechanics as well, but only for the vacuum expectation values (or for a thermal state). For example, one can check it for the 4-point correlation function:
\begin{align}
\langle 0 | \phi_1 \phi_2 \phi_3 \phi_4 | 0 \rangle &= f_1 f_2^* f_3 f_4^* + 2 f_1 f_2 f_3^* f_4^*, \nonum
\langle 0 | \phi_1 \phi_2 | 0 \rangle \langle 0 | \phi_3 \phi_4 | 0 \rangle + \langle 0 | \phi_1 \phi_3 | 0 \rangle \langle 0 | \phi_2 \phi_4 | 0 \rangle + \langle 0 | \phi_1 \phi_4 | 0 \rangle \langle 0 | \phi_2 \phi_3 | 0 \rangle &= f_1 f_2^* f_3 f_4^* + 2 f_1 f_2 f_3^* f_4^*.
\end{align}
It is easy to see that equality does hold. However, this is not the case in the non-vacuum state:
\begin{align}
&\langle 2 | \phi_1 \phi_2 \phi_3 \phi_4 | 2 \rangle = 12 f_1 f_2 f_3^* f_4^* + 9 f_1 f_2^* f_3 f_4^* + 6 f_1 f_2^* f_3^* f_4 + 6 f_1^* f_2 f_3 f_4^* + 4 f_1^* f_2 f_3^* f_4 + 2 f_1^* f_2^* f_3 f_4, \nonum
&\langle 2 | \phi_1 \phi_2 | 2 \rangle \langle 2 | \phi_3 \phi_4 | 2 \rangle + \langle 2 | \phi_1 \phi_3 | 2 \rangle \langle 2 | \phi_2 \phi_4 | 2 \rangle + \langle 2 | \phi_1 \phi_4 | 2 \rangle \langle 2 | \phi_2 \phi_3 | 2 \rangle = \nonum &\phantom{\langle 2 | \phi_1 \phi_2 \phi_3 \phi_4 | 2 \rangle}= 18 f_1 f_2 f_3^* f_4^* + 15 f_1 f_2^* f_3 f_4^* + 12 f_1 f_2^* f_3^* f_4 + 12 f_1^* f_2 f_3 f_4^* + 10 f_1^* f_2 f_3^* f_4 + 8 f_1^* f_2^* f_3 f_4.
\end{align}
Thus, one can build a diagrammatic technique similar to that of the quantum field theory (see section~\ref{sec:diagrams}). However, one cannot apply this technique to an arbitrary state because the Wick's theorem works only for the vacuum or for a thermal state (e.g. see~\cite{Evans}) in quantum mechanics. 

\section{Varying frequency case}
\label{sec:varfreq}

\subsection{Setup of the problem}

In this section we calculate propagators and generalize the result of previous section to the varying frequency case:
\beq
\label{eq:varaction}
S = \int dt \Big[ \frac{1}{2} \dot \phi^2(t) - \frac{\omega^2(t)}{2} \phi^2(t) - \frac{\lambda}{4} \phi^4(t) \Big],
\eeq
where the frequency $\omega(t) \rightarrow \omega_\pm = \text{const}$ as $t \rightarrow \pm \infty$ and the self-interaction $\lambda \phi^4$ turns on adiabatically after $t_0$. Similarly to the constant frequency case we represent the real field $\phi(t)$ as
\beq \phi(t) = a f(t) + a^+ f^*(t), \eeq
where $f(t)$ solves the free equation of motion $\ddot{f} + \omega^2(t) f = 0$. In the WKB approximation modes can be represented as
\beq f(t) = \alpha(t) \exp \Big[ -i \int_{-\infty}^t \omega(t') dt' \Big] + \beta(t) \exp \Big[ i \int_{-\infty}^t \omega(t') dt' \Big]. \eeq
We choose such $f(t)$ that there is only one exponent at the past infinity:
\beq
\label{eq:onewave}
f(t) \simeq \frac{1}{\sqrt{2 \omega_-}} e^{i \omega_- t}, \quad \text{as} \quad t \rightarrow -\infty.
\eeq
This is so-called in-modes. The WKB approximation can be used  when the frequency does not change too fast:
\beq
\label{eq:WKB}
\Big| \frac{d}{dt} \frac{1}{\omega(t)} \Big| \ll 1 \Longrightarrow \frac{\dot \omega(t)}{\omega^2(t)} \ll 1.
\eeq
If this inequality holds for all times $t \in (-\infty, +\infty)$, i.e. frequency changes adiabatically, then we also have single exponent $f(t) \simeq \frac{1}{\sqrt{2 \omega_+}} e^{i \omega_+ t}$, as $t \rightarrow +\infty$. But if the inequality breaks down for some regions of $t$ then we have that:
\beq
\label{eq:fcases}
f(t) = \begin{cases} \frac{1}{\sqrt{2 \omega_-}} e^{i \omega_- t}, \quad \text{as} \quad t \rightarrow -\infty, \\ \frac{\alpha}{\sqrt{2 \omega_+}} e^{i \omega_+ t} + \frac{\beta}{\sqrt{2 \omega_+}} e^{-i \omega_+ t}, \quad \text{as} \quad t \rightarrow +\infty. \end{cases}
\eeq
Here the complex numbers $\alpha$ and $\beta$ satisfy the $|\alpha|^2 - |\beta|^2 = 1$ condition as the consequence of usual commutation relation $[\phi, \pi] = [\phi, \dot{\phi}] = i$ which follows from the time independence of the Wronskian $\dot{f} f^* - f \dot{f}^* = i$. For the beginning we consider the case $\beta \ne 0$, i.e. non-adiabatic case (subsection~\ref{subsec:nonadfreq}). Then we turn to the adiabatic case $\beta = 0$ (subsection~\ref{subsec:adfreq}).

Interaction and evolution operators are defined in the same way as in the previous section (see eq.~\eqref{eq:interaction},~\eqref{eq:evolution} with the appropriate new form of $f(t)$). This time we have $\omega(t) \ne \text{const}$, i.e. there is no time translation invariance and, hence, we do not set $t_0 = 0$. However, below we will show that in the leading expressions the dependence on $t_0$ disappears in the limit $t_0 \rightarrow -\infty$ if we take $f(t)$ as in the equation~\eqref{eq:onewave}.

We want to work in the leading order in powers of $T$, as $T \rightarrow \infty$ and $\tau = \text{const}$ where $T$ and $\tau$ are defined~in~\eqref{eq:times}. Hence, in the leading growing with $T$ expressions we can approximate integrals as follows:
\beq \int_{t_0}^{t_{1/2}} g(t) dt = \int_{t_0}^{T} g(t) dt \pm \frac{\tau}{2} g(T) + O(\tau^2), \quad \int_{t_2}^{t_1} g(t) dt = \tau g(T) + O(\tau^2), \eeq
where the function $g(t)$ changes slowly during the time period $\sim \tau$. Hence, the evolution operator is approximately equal to
\begin{align}
U_1(t_{1/2}, t_0) &= -i \int_{t_0}^{t_{1/2}} V(\eta) d\eta = -i \int_{t_0}^{T} V(\eta) d\eta + O(T^0), \nonum
U_1(t_1, t_2) &= -i \int_{t_2}^{t_1} V(\eta) d\eta = -i \tau V(T) + O(\tau^2) = O(T^0), \nonum
U_2(t_1, t_2) &= -i \int_{t_2}^{t_1} d\eta V(\eta) U_1(\eta, t_2) = -i \tau V(T) U_1(T, t_2) + O(\tau^2) = O(T^0), \nonum
U_2(t_{1/2}, t_0) &= -i \int_{t_0}^{t_{1/2}} d\eta V(\eta) U_1(\eta, t_0) = - \int_{t_0}^T d\eta V(\eta) \int_{t_0}^\eta d\xi V(\xi) \mp \frac{\tau}{2} V(T) \int_{t_0}^T d\eta V(\eta) + O(T^0), \nonum
U_1(t_0, t_{1/2}) &= -i \int_{t_{1/2}}^{t_0} d\eta V(\eta) U_1(\eta, t_1) = - \int_{t_0}^T d\eta V(\eta) \int_{\eta}^T d\xi V(\xi) \mp \frac{\tau}{2} \int_{t_0}^T d\eta V(\eta) V(T) + O(T^0).
\end{align}
To single out leading expressions we keep only non-oscillating terms in $V(t)$, as $t = T \rightarrow +\infty$. These are:
\begin{gather}
(2 \omega_+)^2 \cdot f^4(T) \simeq 6 \alpha^2 \beta^2, \quad (2 \omega_+)^2 \cdot f^{*4}(T) \simeq 6 \alpha^{*2} \beta^{*2}, \nonum
(2 \omega_+)^2 \cdot |f(T)|^2 f^2(T) \simeq 3 \alpha \beta \big( |\alpha|^2 + |\beta|^2 \big), \quad (2 \omega_+)^2 \cdot |f(T)|^2 f^{*2}(T) \simeq 3 \alpha^* \beta^* \big( |\alpha|^2 + |\beta|^2 \big), \nonum
\label{eq:non-oscillating-parts}
(2 \omega_+)^2 \cdot |f(T)|^4 = |\alpha|^4 + 4 |\alpha|^2 |\beta|^2 + |\beta|^4.
\end{gather}
It is easy to see that the only non-oscillating term from $V(t)$, as $ t = t_0 \rightarrow -\infty $, is as follows:
\beq (2 \omega_-)^2 \cdot |f(t)|^4 = 1. \eeq
When we perform integrations, we should care only about such non-oscillating expressions because they give the leading contribution to relevant expressions in the limit $T \rightarrow \infty$, while $\tau = \text{const}$.

Finally, for the calculations below it is important that the expectation values of the operators~\eqref{eq:AB} are real and have the following properties:
\beq \langle n | a^{+2} B | n \rangle = \langle n | a^{+2} B | n \rangle^* = \langle n | B^+ a^2 | n \rangle, \quad \langle n | a^2 B^+ | n \rangle = \langle n | B a^{+2} | n \rangle. \eeq

\subsection{Non-adiabatically varying frequency case}
\label{subsec:nonadfreq}

Let us calculate the two-point correlation function in the $n$-th eigen-state of the free Hamiltonian $H_0 | n \rangle = n | n \rangle$:

\beq \langle \phi(t_1) \phi(t_2) \rangle = \lim_{\begin{subarray}{c} t_0 \rightarrow -\infty \\ T \rightarrow +\infty \end{subarray}} \langle n| U^+(t_1, t_0) \phi(t_1) U^+(t, t_1) U(t, t_2) \phi(t_2) U(t_2, t_0) |n \rangle. \eeq
It is convenient to split this expression into the terms $C_0$, $C_1$, $C_2$ and denote $\phi(t_a) \equiv \phi_a$, $C(t_a, t_b) \equiv C(a, b)$ as in the constant frequency case~\eqref{eq:correlator},~\eqref{eq:correlators}.

At the zeroth order in $\lambda$:
\beq C_0(1, 2) = \langle n | \phi_1 \phi_2 | n \rangle = (n + 1)f_1 f_2^* + n f_1^* f_2. \eeq
In the first order in $\lambda$:
\begin{align}
\label{eq:1st}
C_1(1, 2) &= \frac{\lambda}{4} \Big\langle n \Big| \Big( i \int_{t_0}^T d\eta V(\eta) \Big) \cdot \Big( f_1 f_2 a^2 + f_1^* f_2^* a^{+2} + f_1^* f_2 a^+ a + f_1 f_2^* a a^+ \Big) + \nonum &+ \Big( f_1 f_2 a^2 + f_1^* f_2^* a^{+2} + f_1^* f_2 a^+ a + f_1 f_2^* a a^+ \Big) \cdot \Big( -i \int_{t_0}^T d\eta V(\eta) \Big) \Big| n \Big\rangle + O(T^0) = \nonum
&= 6 \lambda i \cdot \Big( n^2 + n + \frac{1}{2} \Big) \cdot \int_{t_0}^T d\eta |f(\eta)|^2 \Big( f^2(\eta) f_1^* f_2^* - f^{*2}(\eta) f_1 f_2 \Big) + O(T^0).
\end{align}
Keeping non-oscillating integrands at the future and past infinities, we obtain:
\beq C_1(1, 2) \simeq -18 i \cdot \frac{\lambda T}{(2 \omega_+)^2} \cdot \Big( n^2 + n + \frac{1}{2} \Big) \cdot \big( |\alpha|^2 + |\beta|^2 \big) \Big( \alpha^* \beta^* f_1 f_2 - \alpha \beta f_1^* f_2^* \Big) + O(T^0). \eeq
Let us stress here that there is no growth as $t_0 \rightarrow -\infty$ since the integrands under the time integrals rapidly oscillate in this limit. In other words, the dependence on $t_0$ disappears from the leading order expressions. This is the property of the in-modes, because $f_{in}(t) \sim e^{i \omega t}$, as $t \rightarrow -\infty$. 

It is convenient to split the second order expression as $C_2(1, 2) = C_2^1(1, 2) + C_2^2(1, 2)$, where
\begin{align}
\label{eq:var2nd}
C_2^1(1, 2) &= \Big\langle \int_{t_0}^T d\eta V(\eta) \phi_1 \phi_2 \int_{t_0}^T d\xi V(\xi) \Big\rangle + \nonum &+ \tau \int_{t_0}^T d\eta \Big\langle V(\eta) \phi_1 V(T) \phi_2 - \phi_1 V(T) \phi_2 V(\eta) \Big\rangle + \nonum &+ \frac{\tau}{2} \int_{t_0}^T d\eta \Big\langle V(T) \phi_1 \phi_2 V(\eta) - V(\eta) \phi_1 \phi_2 V(T) \Big\rangle + O(T^0), \nonum
C_2^2(1, 2) &= - \Big\langle \int_{t_0}^T d\eta V(\eta) \int_\eta^T d\xi V(\xi) \phi_1 \phi_2 \Big\rangle - \Big\langle \phi_1 \phi_2 \int_{t_0}^T d\eta V(\eta) \int_{t_0}^\eta d\xi V(\xi) \Big\rangle + \nonum &+ \frac{\tau}{2} \int_{t_0}^T d\eta \Big\langle \phi_1 \phi_2 V(T) V(\eta) - V(\eta) V(T) \phi_1 \phi_2 \Big\rangle + O(T^0).
\end{align}
Calculating $C_2^1(1, 2)$ and $C_2^2(1, 2)$ separately, keeping the leading order $\sim T^2$ expressions and combining them again one finds that in an arbitrary eigen-state of the free Hamiltonian:
\begin{align}
\frac{1}{\lambda^2} C_2(1, 2) &= 
f_1 f_2^* \cdot (4 n^3 + 6 n^2 + 14 n + 6) \cdot \int_{t_0}^T d\eta f^4(\eta) \int_{t_0}^T d\xi f^{*4}(\xi) + \nonum
&+ f_1 f_2^* \cdot (16 n^3 + 24 n^2 + 26 n + 9) \cdot \int_{t_0}^T d\eta f^2(\eta) |f(\eta)|^2 \int_{t_0}^T d\xi f^{*2}(\xi) |f(\xi)|^2 + \nonum
&+ f_1 f_2 \cdot (16 n^3 + 24 n^2 + 26 n + 9) \cdot \int_{t_0}^T d\eta f^2(\eta) |f(\eta)|^2 \int_\eta^T d\xi f^{*4}(\xi) - \nonum
&- f_1 f_2 \cdot (8 n^3 + 12 n^2 + 28 n + 12) \cdot \int_{t_0}^T d\eta f^{*4}(\eta) \int_\eta^T d\xi f^2(\xi) |f(\xi)|^2 - \nonum
&- f_1 f_2 \cdot (48 n^3 + 72 n^2 + 78 n + 27) \cdot \int_{t_0}^T d\eta f^{*2}(\eta) |f(\xi)|^2 \int_\eta^T d\xi |f(\xi)|^4 + h.c. + O(T).
\end{align}
Here we have used that
\beq
\label{eq:change-limits}
\int_{t_0}^T d\eta \, g(\eta) \int_{t_0}^\eta d\xi \, h(\xi) = \int_{t_0}^T d\xi \, h(\xi) \int_\eta^T d\eta \, g(\eta).
\eeq
Keeping non-oscillating integrands in the limit as $T \rightarrow +\infty$, $t_0 \rightarrow -\infty$, $\tau = \text{const}$, we obtain in the leading order that:
\begin{align}
\label{eq:C2-nonad-true}
C_2(1, 2) &= \Bigg[ \Bigg( 4 \Big( n + \frac{1}{2} \Big)^3 - \Big( n + \frac{1}{2} \Big) \Bigg) \cdot 36 \big( 1 + 5 |\alpha|^2 |\beta|^2 \big) + \Big( n + \frac{1}{2} \Big) \cdot 54 \big( 3 + 20 |\alpha|^2 |\beta|^2 \big) \Bigg] \cdot \nonum &\cdot \frac{(\lambda T)^2}{(2 \omega_+)^4} \cdot \Big[ |\alpha|^2 |\beta|^2 \Big( f_1 f_2^* + f_1^* f_2 \Big) - \frac{1}{2} \big( |\alpha|^2 + |\beta|^2 \big) \Big( \alpha \beta f_1^* f_2^* + \alpha^* \beta^* f_1 f_2 \Big) \Big] + O(T),
\end{align}
where we used the fact that $|\alpha|^2 - |\beta|^2 = 1$. Note that after the substitution of the modes~\eqref{eq:fcases} one obtains:
\beq |\alpha|^2 |\beta|^2 f_1 f_2^* - \frac{1}{2} \big( |\alpha|^2 + |\beta|^2 \big) \cdot \alpha \beta f_1^* f_2^* + h.c. = -\frac{1}{2} \big( |\alpha|^2 + |\beta|^2 \big) \Big( \alpha \beta^* e^{2 i \omega T} + \alpha^* \beta e^{-2 i \omega T} \Big), \eeq
so the expression~\eqref{eq:C2-nonad-true} does not depend on $\tau$. The same expressions in the vacuum state $H_0 | 0 \rangle = \frac{1}{2} | 0 \rangle$ are as follows:
\begin{align}
\label{eq:2nd}
\frac{1}{3\lambda^2} C_2(1, 2) &= 
f_1 f_2^* \cdot \Bigg( 2 \int_{t_0}^T d\eta f^4(\eta) \int_{t_0}^T d\xi f^{*4}(\xi) + 3 \int_{t_0}^T d\eta f^2(\eta) |f(\eta)|^2 \int_{t_0}^T d\xi f^{*2}(\xi) |f(\xi)|^2 \Bigg) + \nonum
&- f_1 f_2 \cdot \Bigg( 4 \int_{t_0}^T d\eta f^{*4}(\eta) \int_\eta^T d\xi f^2(\xi) |f(\xi)|^2 - 3 \int_{t_0}^T d\eta f^2(\eta) |f(\eta)|^2 \int_\eta^T d\xi f^{*4}(\xi) \Bigg) - \nonum
&- f_1 f_2 \cdot 9 \int_{t_0}^T d\eta f^{*2}(\eta) |f(\xi)|^2 \int_\eta^T d\xi |f(\xi)|^4 + h.c. + O(T),
\end{align}
and
\begin{align}
\label{eq:2nd-in-modes}
C_2(1, 2) &= \frac{(\lambda T)^2}{(2 \omega_+)^4} \cdot 27 \big( 3|\alpha|^4 + 3|\beta|^4 + 14 |\alpha|^2 |\beta|^2 \big) \cdot \nonum &\cdot \Big[ |\alpha|^2 |\beta|^2 \Big( f_1 f_2^* + f_1^* f_2 \Big) - \frac{1}{2} \big( |\alpha|^2 + |\beta|^2 \big) \Big( \alpha \beta f_1^* f_2^* + \alpha^* \beta^* f_1 f_2 \Big) \Big] + O(T).
\end{align}
Let us stress a few important properties of these expressions. First, the dependence on $t_0$ disappear as in the first order case~\eqref{eq:1st}. Second, the leading order is symmetric under the change $t_1 \leftrightarrow t_2$. This means that the leading order of the Keldysh propagator is exactly equal to $C_2 \sim (\lambda T)^2$ whereas retarded and advanced propagators are $O(\lambda^2 T)$, i.e. they are suppressed by higher powers of $\lambda$ as $\lambda \rightarrow 0$, $T \rightarrow \infty$ and $\lambda T = \text{const}$.

Finally, in the vacuum state $H_0 | 0 \rangle = \frac{1}{2} | 0 \rangle$ we obtain that:
\begin{align}
\label{eq:direct-propagators}
D_0^K(t_1, t_2) &= \frac{1}{2} \Big( f_1 f_2^* + f_1^* f_2 \Big), \nonum
D_1^K(t_1, t_2) &\simeq -9 i \cdot \frac{\lambda T}{(2 \omega_+)^2} \cdot (|\alpha|^2 + |\beta|^2) \cdot \Big( \alpha^* \beta^* f_1 f_2 - \alpha \beta f_1^* f_2^* \Big) + O(T^0), \nonum
D_2^K(t_1, t_2) &= \frac{(\lambda T)^2}{(2 \omega_+)^4} \cdot 27 \big( 3|\alpha|^4 + 3|\beta|^4 + 14 |\alpha|^2 |\beta|^2 \big) \cdot \nonum &\cdot \Big[ |\alpha|^2 |\beta|^2 \Big( f_1 f_2^* + f_1^* f_2 \Big) - \frac{1}{2} \big( |\alpha|^2 + |\beta|^2 \big) \Big( \alpha \beta f_1^* f_2^* + \alpha^* \beta^* f_1 f_2 \Big) \Big] + O(T), \nonum
D_0^R(t_1, t_2) &= \theta (t_1 - t_2) \Big( f_1 f_2^* - f_1^* f_2 \Big), \nonum
D_0^A(t_1, t_2) &= -\theta (t_2 - t_1) \Big( f_1 f_2^* - f_1^* f_2 \Big), \nonum
D_1^R(t_1, t_2) &= D_1^A(t_1, t_2) = O(T^0), \nonum
D_2^R(t_1, t_2) &= D_2^A(t_1, t_2) = O(T).
\end{align}
Let us stress here that the Keldysh propagator is real, while R/A propagators are imaginary in all orders as should be expected.

In conclusion, we have obtained the secular growth of the loop corrections to the Keldysh propagator in both the first and the second orders in $\lambda$: they grow indefinitely in the limit $T \rightarrow \infty$, $\tau = \text{const}$. Looking on the form of~\eqref{eq:direct-propagators} and comparing it with~\eqref{eq:nk} one can see that level population $n = \langle a^+ a \rangle$ and anomalous quantum average $\kappa = \langle a a \rangle$ also grow indefinitely in such a limit. At the same time, corrections to the retarded and advanced propagators are suppressed by higher powers of $\lambda$ in the same limit.

\subsection{Adiabatically varying frequency case}
\label{subsec:adfreq}

Let us consider now the case of adiabatically varying frequency. This corresponds to the case of $\beta = 0$ in the eq.~\eqref{eq:fcases}, i.e. the mode functions possess the following behavior:
\beq
\label{eq:fcases-ad}
f(t) = \begin{cases} \frac{1}{\sqrt{2 \omega_-}} e^{i \omega_- t}, \quad \text{as} \quad t \rightarrow -\infty, \\ \frac{1}{\sqrt{2 \omega_+}} e^{i \omega_+ t}, \quad \text{as} \quad t \rightarrow +\infty. \end{cases}
\eeq
In this case one can use the general expressions obtained in the previous subsection~\eqref{eq:1st},~\eqref{eq:2nd} replacing modes~\eqref{eq:fcases} with those from the eq.~\eqref{eq:fcases-ad}. From there one can see that the first order correction in $\lambda$ does not grow with time:
\beq C_1(1, 2) = -6 \lambda i \cdot \Big( n^2 + n + \frac{1}{2} \Big) \cdot \int_{t_0}^T d\eta |f(\eta)|^2 \Big( f^2(\eta) f_1^* f_2^* - f^{*2}(\eta) f_1 f_2 \Big) + O(T^0) = O(T^0). \eeq
Hence, neither Keldysh nor R/A propagators grow as $T \rightarrow \infty$: $D_1^K(1, 2) \sim D_1^R(1, 2) \sim O(T^0)$.

The second order correction in $\lambda$ is more sophisticated. This time we should take into account the following term which in contrast to the non-adiabatic case is not negligible:
\beq
\label{eq:2nd-non-oscillating-ad}
\int_{t_0}^T d\eta f^2(\eta) |f(\eta)|^2 \int_\eta^T d\xi |f(\xi)|^4 \simeq \int_{t_0}^T d\eta f^2(\eta) |f(\eta)|^2 \frac{T - \eta}{(2 \omega(\eta))^2} \simeq \Big( \frac{T}{(2 \omega_+)^2} - \frac{t_0}{(2 \omega_-)^2} \Big) \frac{i e^{2 i \omega_- t_0}}{(2 \omega_-)^3}.
\eeq
Here we perform integration by parts assuming that due to adiabaticity
\beq d\Big( \frac{e^{i \omega(t) t}}{i \omega(t)} \Big) = e^{i \omega(t) t} + e^{i \omega(t) t} \cdot \frac{t \omega'(t)}{\omega(t)} \simeq e^{i \omega(t) t}. \eeq
The dependence on the time $t_0$ reappears here due to the non-oscillating integrand $|f(\xi)|^4 = \big( 2 \omega(t) \big)^{-2}$ which gives the factor $T - \eta$, although the first integrand $f^2(\eta) |f(\eta)|^2$ still oscillates. In the non-adiabatic case $\beta \ne 0$ these integrals were suppressed in comparision with the leading order expressions (see eq.~\eqref{eq:2nd-in-modes}). But in the case $\beta = 0$ the leading order is $\sim T$ due to the behaviour of the modes~\eqref{eq:fcases-ad}: all the integrals from the expression~\eqref{eq:2nd} except~\eqref{eq:2nd-non-oscillating-ad} contain only oscillating integrands, and hence they can be omitted.

Also we should take into account terms with one integration which can give the $\sim T \tau$ dependence in~\eqref{eq:var2nd}. However, they cancel each other due to the properties of the operator $A$:
\begin{align}
C_2(1, 2) \Big|_\tau &= \tau \int_{t_0}^T d\eta \Big\langle V(\eta) \phi_1 V(T) \phi_2 - \phi_1 V(T) \phi_2 V(\eta) \Big\rangle + \nonum &+ \frac{\tau}{2} \int_{t_0}^T d\eta \Big\langle V(T) \phi_1 \phi_2 V(\eta) - V(\eta) \phi_1 \phi_2 V(T) \Big\rangle + \nonum &+ \frac{\tau}{2} \int_{t_0}^T d\eta \Big\langle \phi_1 \phi_2 V(T) V(\eta) - V(\eta) V(T) \phi_1 \phi_2 \Big\rangle + O(T^0) = \nonum
&= \tau \cdot \Big( \frac{T}{(2 \omega_+)^2} - \frac{t_0}{(2 \omega_-)^2} \Big) \cdot \langle 3 A \rangle \cdot \Big( \phi_1 V(T) \phi_2 - \phi_1 V(T) \phi_2 + \nonum &+ \frac{1}{2} V(T) \phi_1 \phi_2 - \frac{1}{2} \phi_1 \phi_2 V(T) + \frac{1}{2} \phi_1 \phi_2 V(T) - \frac{1}{2} V(T) \phi_1 \phi_2 \Big) + O(T^0) = \nonum &= O(T^0).
\end{align}
Let us notice that such a cancellation occurs not only in the vacuum state, but in an arbitrary eigen-state of the free Hamiltonian as well. In the vacuum state one obtains:
\begin{align}
\label{eq:C2-ad}
\frac{1}{\lambda^2} C_2(1, 2) &\simeq - f_1 f_2 \cdot 27 \int_{t_0}^T d\eta f^{*2}(\eta) |f(\xi)|^2 \int_\eta^T d\xi |f(\xi)|^4 + h.c. + O(\tau) = \nonum
&= \frac{27 i}{(2 \omega_-)^3} \Big( \frac{T}{(2 \omega_+)^2} - \frac{t_0}{(2 \omega_-)^2} \Big) \Big( e^{2 i \omega_- t_0} f_1^* f_2^* - e^{-2 i \omega_- t_0} f_1 f_2 \Big) + O(T^0).
\end{align}
Corrections to the R/A propagators do not grow $D_2^R(1, 2) = O(T^0)$ due to the symmetry of $C_2(1, 2)$ under the change $t_1 \leftrightarrow t_2$. At the same time, the corrections to the Keldysh propagator are the same as to the Wightman correlation function $D_2^K(1, 2) = C_2(1, 2) = O(T)$. Thus, in the case of adiabatic time dependence of $\omega(t)$ we obtain the secular growth again. However, it is slower than in the non-adiabatically varying frequency case and can be removed by the redefinition of the vacuum state as in the constant frequency case (see subsection~\ref{subsec:causes}).

It is obvious that the expression~\eqref{eq:C2-ad} turns into the expression~\eqref{eq:C2-const} from the subsection~\ref{subsec:constfreq} if one sets $\omega_- = \omega_+ = \omega$ and $t_0 = 0$, as it should be expected.

\section{Diagrammatic technique}
\label{sec:diagrams}

\subsection{Brief introduction into the diagrammatic technique}
\label{subsec:brief}

In this section we introduce the diagrammatic technique which allows one to calculate the loop corrections to the propagators. This technique gives right combinatoric factors and reproduces the result of direct calculations performed above in the sections~\ref{sec:constfreq}~and~\ref{sec:varfreq}. However, we emphasize that this technique works only in the vacuum state bacause it uses Wick's theorem (see subsection~\ref{subsec:Wick}). We assume that the frequency $\omega(t)$ changes non-adiabatically, i.e. $\beta \ne 0$.

First, we consider the following partition function~\cite{Kamenev, Berges, Rammer, Calzetta, Millington}:
\begin{align}
Z_0[\phi(t)] &= \langle 0 | U_{\mathcal{C}} | 0 \rangle = \int \mathcal{D}\phi e^{i S_0[\phi(t)]}, \\
\label{eq:varaction-path}
S_0[\phi(t)] &= S_0[\phi_+, \phi_-] = \int_{\mathcal{C}} dt \Big[ \frac{1}{2} \phi^2 - \frac{\omega^2}{2} \phi^2 \Big] = \int_{t_0}^{+\infty} dt \Big[ \frac{1}{2} \dot{\phi_{+}^2} - \frac{\omega^2}{2} \phi_{+}^2 -\frac{1}{2} \dot{\phi_{-}^2} + \frac{\omega^2}{2} \phi_{-}^2 \Big] = \nonum &= \frac{i}{2} \int dt \vec{\phi}^{\,T}(t) \hat{G}_0^{-1} \vec{\phi}(t),
\end{align}
where $\mathcal{D}\phi$ denotes the integration over all real scalar functions $\phi$, $\mathcal{C}$ is the Keldysh time contour~\cite{Keldysh}, $U_{\mathcal{C}}$ is evolution operator time ordered along the contour $\mathcal{C}$ and '$\pm$' subscripts correspond to the forward and backward parts of this contour. We have introduced vector $\vec{\phi}$ and inversed operator $\hat{G}_0^{-1}$ for short:
\beq
\label{eq:inversedG}
\vec{\phi} = \Big(\begin{array}{c} \phi_+ \\ \phi_- \end{array}\Big), \quad \hat{G}_0^{-1} = -i \Big(\begin{array}{cc} \partial_t^2 + \omega^2(t) & 0 \\ 0 & -\partial_t^2 - \omega^2(t) \end{array}\Big).
\eeq
Dividing contour $\mathcal{C}$ into $2N$ equal time intervals and discretizing the action we obtain:
\begin{align}
Z_d &= \int \prod_{j = 1}^{2 N} \frac{d \phi_j}{\sqrt{2 \pi \delta t}} \exp\Big( i S_d[\phi] \Big) = \int \prod_{j = 1}^{2 N} \frac{d \phi_j}{\sqrt{2 \pi \delta t}} \exp\Big( -\frac{1}{2} \sum_{k, m = 1}^{2 N} \phi_k G_{km}^{-1} \phi_m \Big), \\
S_d[\phi] &= \sum_{j = 1}^{2 N} \Big( \frac{1}{\delta t_j} - \frac{\omega_j^2 \delta t_j}{2} \Big) \phi_j^2 - \sum_{j = 2}^{2 N - 1} \frac{1}{2 \delta t_j} ( \phi_j \phi_{j - 1} + \phi_j \phi_{j + 1} ) - \frac{1}{2 \delta t_1} \phi_1 \phi_2 - \frac{1}{2 \delta t_{2 N}} \phi_{2 N} \phi_{2 N - 1} = \nonum &= \sum \delta t_j \Big[ \frac{(\phi_j - \phi_{j - 1})^2}{2 \delta t_j^2} - \frac{\omega_j^2}{2} \phi_j^2 \Big] + \cdots,
\end{align}
where $\delta t_j = \pm \delta t$ is the length of the $j$-th time interval on the forward and backward parts of contour $\mathcal{C}$ correspondingly and $\phi_j = \phi(t_j)$, $\omega_j = \omega(t_j)$. In the continuum limit $N \rightarrow \infty$ `tail' parts containing $\phi_1$ and $\phi_{2N}$ are negligible and discrete action $S_d[\phi]$ turns into the action $S_0[\phi]$; one can find more details in the book~\cite{Kamenev}.

It is easy to obtain two-field correlation functions using this discrete partition function and calculating the following Gaussian integral:
\beq Z_d[J] = \int \prod_{j = 1}^{N} \frac{d\phi_j}{\sqrt{2 \pi}} \exp\Big( -\frac{1}{2} \sum_{k, m = 1}^{N} \phi_k A_{km} \phi_m + \sum_{k = 1}^{N} \phi_k J_k \Big) = \frac{1}{\sqrt{\det A}} e^{\frac{1}{2} \sum J_k A_{km}^{-1} J_m }. \eeq
Differentiating this expression and setting $J = 0$ we obtain the corresponding rules for the discretized fields:
\begin{align}
\label{eq:Wick-rules}
\langle \phi_a \phi_b \rangle &= \frac{1}{Z_d[0]} \frac{\delta^2 Z[J]}{\delta J_a \delta J_b} \Bigg\vert_{J = 0} = A_{ab}^{-1}, \nonum
\langle \phi_a \phi_b \phi_c \phi_d \rangle &= A_{ab}^{-1} A_{cd}^{-1} + A_{ac}^{-1} A_{bd}^{-1} + A_{ad}^{-1} A_{bc}^{-1} = \nonum &= \langle \phi_a \phi_b \rangle \langle \phi_c \phi_d \rangle + \langle \phi_a \phi_c \rangle \langle \phi_b \phi_d \rangle + \langle \phi_a \phi_d \rangle \langle \phi_b \phi_c \rangle,
\end{align}
and so on. These rules are similar to the Wick's rules from quantum field theory (see subsection~\ref{subsec:Wick} for details).

Returning to the continuum limit and distinguishing fields corresponding to the forward part ($\phi_j \rightarrow \phi_+$, $j = 1, \cdots, N$) and backward part ($\phi_j \rightarrow \phi_-$, $j = 2 N, \cdots, N + 1$) we obtain the following two-point correlation functions if the average is taken with respect to the ground state:
\begin{align}
\langle \phi_+(t_1) \phi_-(t_2) \rangle_0 &= G_0^{+-}(t_1, t_2) = f_1 f_2^*, \nonum
\langle \phi_-(t_1) \phi_+(t_2) \rangle_0 &= G_0^{-+}(t_1, t_2) = f_1^* f_2, \nonum
\langle \phi_+(t_1) \phi_+(t_2) \rangle_0 &= G_0^{++}(t_1, t_2) = \theta(t_1 - t_2) f_1^* f_2 + \theta(t_2 - t_1) f_1 f_2^*, \nonum
\langle \phi_-(t_1) \phi_-(t_2) \rangle_0 &= G_0^{--}(t_1, t_2) = \theta(t_1 - t_2) f_1 f_2^* + \theta(t_2 - t_1) f_1^* f_2,
\end{align}
where $f_1 = f(t_1)$ and $f_2 = f(t_2)$ are the modes from the previous sections~\eqref{eq:simplemode},~\eqref{eq:fcases} or~\eqref{eq:fcases-ad}. So one can calculate correlation functions of an arbitrary number of the forward and backward fields using these propagators and Wick's theorem~\eqref{eq:Wick-rules} (i.e. splitting the average into the sum of all possible pair contractions). Here we denoted by angular brackets the averaging in the following sense:
\beq \langle \phi_\pm(t) \phi_\pm(t') \rangle_0 \equiv \int \mathcal{D} \phi \, \phi_\pm(t) \phi_\pm(t') e^{i S_0[\phi]}. \eeq
Note that $Z_0[\phi(t)] = \int \mathcal{D} \phi e^{i S_0[\phi]} = 1$ if one considers closed Keldysh time contour $\mathcal{C}$ in $S_0$.

Let us emphasize that we can not separate the integrals over $\phi_+$ and $\phi_-$ because of non-vanishing cross-correlation between these fields. It is easy to check that operator $\hat{G}_0$ composed of these components is the inverse of $\hat{G}_0^{-1}$ from the partition function~\eqref{eq:inversedG}:
\beq
\label{eq:notinversedG}
\hat{G}_0(t, t') = \Bigg(\begin{array}{cc} G_0^{++}(t, t') & G_0^{+-}(t, t') \\ G_0^{-+}(t, t') & G_0^{--}(t, t') \end{array}\Bigg), \quad \hat{G}_0^{-1} \hat{G}_0 = \hat{G}_0 \hat{G}_0^{-1} = \hat{1}.
\eeq
However, not all of four propagators are independent:
\beq G_0^{++} + G_0^{--} = G_0^{+-} + G_0^{-+}, \eeq
so it is convenient to make the Keldysh rotation:
\beq
\label{eq:keldysh-rotation}
\Big(\begin{array}{c} \phi_{cl} \\ \phi_{q}\end{array}\Big) = \hat{R} \Big(\begin{array}{c} \phi_{+} \\ \phi_{-}\end{array}\Big), \quad {\rm where} \quad \hat{R} = \hat{R}^{-1} = \frac{1}{\sqrt{2}} \Big(\begin{array}{cc} 1 & 1 \\ 1 & -1 \end{array}\Big), \quad \det \hat{R} = 1.
\eeq 
After the rotation the action changes in the following way:
\beq S_0[\phi(t)] = \int_{t_0}^{+\infty} dt \Big[ \dot{\phi}_{cl} \dot{\phi}_q - \omega^2 \phi_{cl} \phi_q  \Big], \eeq
and operator $\hat{G}_0$ acquires the form of:
\beq \hat{G}_0(t, t') = \Big(\begin{array}{cc} 2 G_0^K(t, t') & G_0^R(t, t') \\ G_0^A(t, t') & 0 \end{array}\Big). \eeq
One can see that these functions are exactly equal to the Keldysh and R/A free propagators~\eqref{eq:direct-propagators}. So in what follows we assume $G(t, t') \equiv D(t, t')$.

\begin{figure}[ht]
\center{\includegraphics[scale=0.5]{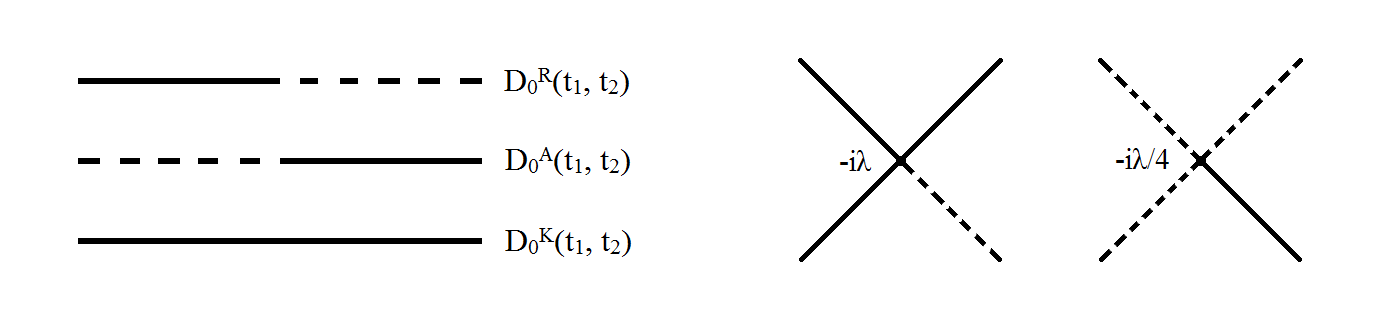}}\caption{The solid line corresponds to the $\phi_{cl}$, the dashed one --- to the $\phi_q$}\label{fig:technique}
\end{figure}

Now let us add $\lambda \phi^4$ self-interaction into the theory:
\begin{align}
\label{eq:path-interaction}
Z[\phi(t)] &= \int \mathcal{D}\phi e^{i S[\phi(t)]}, \nonum
S[\phi(t)] &= \int_{\mathcal{C}} dt \Big[ \frac{1}{2} \phi^2 - \frac{\omega^2}{2} \phi^2 - \frac{\lambda}{4} \phi^4 \Big] = \int_{t_0}^{+\infty} dt \Big[ \frac{1}{2} \dot{\phi_{+}^2} - \frac{\omega^2}{2} \phi_{+}^2 - \frac{\lambda}{4} \phi_+^4 - (\phi_+ \leftrightarrow \phi_-) \Big].
\end{align}
Decomposing the exponential function and using Wick's theorem~\eqref{eq:Wick-rules} one can calculate corrections to the propagators. For example, the exact propagator $D^{++}$ is equal to:
\begin{align}
\label{eq:path-decomposition}
D^{++}(t_1, t_2) &= \langle \phi_1^+ \phi_2^+ \rangle \equiv \int \mathcal{D}\phi \, \phi_+(t_1) \phi_+(t_2) e^{i S[\phi]} = \nonum &= \langle \phi_1^+ \phi_2^+ \rangle_0 - \frac{i \lambda}{4} \int dt_3 \, \Big[ \langle \phi_1^+ \phi_3^+ \phi_3^+ \phi_3^+ \phi_3^+ \phi_2^+ \rangle_0 - \langle \phi_1^+ \phi_3^- \phi_3^- \phi_3^- \phi_3^- \phi_2^+ \rangle_0 \Big] + \cdots = \nonum &= D_0^{++}(t_1, t_2) - D_0^{++}(t_1, t_2) \cdot \frac{3 i \lambda}{4} \int dt_3 \Big[ \big( D_0^{++}(t_3, t_3) \big)^2 - \big( D_0^{--}(t_3, t_3) \big)^2 \Big] - \nonum &- 3 i \lambda \int dt_3 \Big[ D_0^{++}(t_1, t_3) D_0^{++}(t_3, t_3) D_0^{++}(t_3, t_2) - D_0^{+-}(t_1, t_3) D_0^{--}(t_3, t_3) D_0^{-+}(t_3, t_2) \Big] + \nonum &+ \cdots.
\end{align}
Here we denoted for short $\phi_a^+ \equiv \phi_+(t_a)$, $\phi_a^- \equiv \phi_-(t_a)$. Also we contracted the fields and took into account that some contractions are equal to each other which gave the numerical coefficients 3 and 12. For example, the following expressions are equal:
\beq
\label{eq:contractions}
\langle \underbrace{\phi_1^+ \phi_3^-} \overbrace{\phi_3^- \phi_3^-} \underbrace{\phi_3^- \phi_2^+} \rangle_0 = \langle \lefteqn{\underbrace{\phantom{\phi_1^+ \phi_3^- \phi_3^-}}} \phi_1^+ \overbrace{\phi_3^- \phi_3^- \phi_3^-} \underbrace{\phi_3^- \phi_2^+} \rangle_0.
\eeq
One can build a diagrammatic technique using this decomposition. However, it is more convenient to work after the Keldysh rotation~\eqref{eq:keldysh-rotation}:
\beq S[\phi(t)] = \int_{t_0}^{+\infty} dt \Big[ \dot{\phi}_{cl} \dot{\phi}_q - \omega^2 \phi_{cl} \phi_q  - \lambda \Big( \phi_{cl}^3 \phi_{q} + \frac{1}{4} \phi_{cl} \phi_{q}^3 \Big) \Big]. \eeq
Decomposing the exponential function in~\eqref{eq:path-interaction} and contracting the fields in the way similar to~\eqref{eq:path-decomposition} we obtain the diagrammatic technique which resembles the diagrammatic technique from the quantum field theory (Fig.~\ref{fig:technique}). One can find the applications of such a technique in the quantum field theory in~\cite{Berges, Rammer, Calzetta, Millington, Mueller, Akhmedov:2013-phi3, Akhmedov:2013-phi4, Polyakov:2012}, etc. 

\begin{figure}[ht]
\center{\includegraphics[scale=0.75]{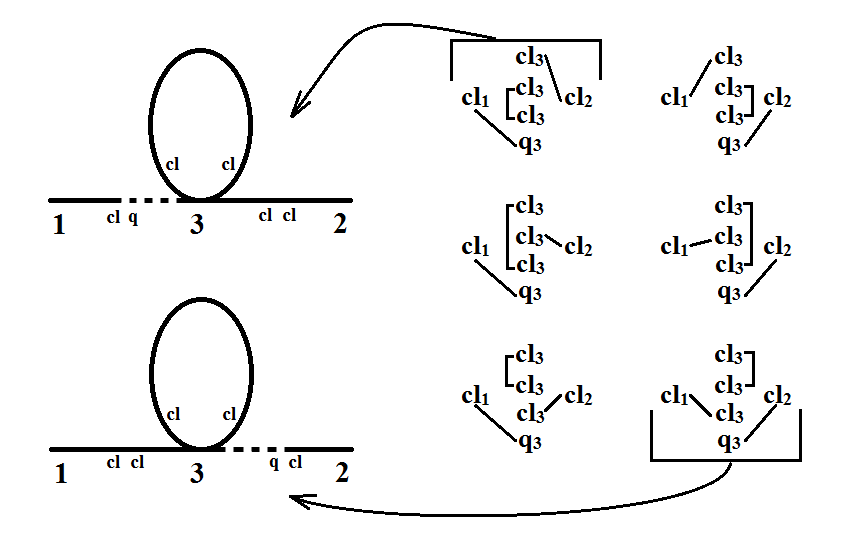}}\caption{`Tadpole' corrections to the Keldysh propagator (on the left side) and corresponding expressions (on the right side)}\label{fig:new-technique-1}
\end{figure}

However, there are some subtleties which are related to the combinatoric factors. In fact, the diagrams of the form (Fig.~\ref{fig:technique}) do not restore these factors fully correctly, although they give the correct relative ratios. Let us consider the first order correction to the Keldysh propagator to show this. At first sight, this correction is given by two `tadpole' one-loop diagrams. However, if one takes into account the order of contracting fields one sees that both these diagrams correspond to three equal expressions (see Fig.~\ref{fig:new-technique-1}). This means that the result obtained with standard diagrams (Fig.~\ref{fig:technique}) must be multiplied by the numerical factor 3. Such factors play an important role when we combine the expressions corresponding to the different types of diagrams (e.g. see subsection~\ref{subsec:diagrams-two}).

\begin{figure}[ht]
\center{\includegraphics[scale=0.75]{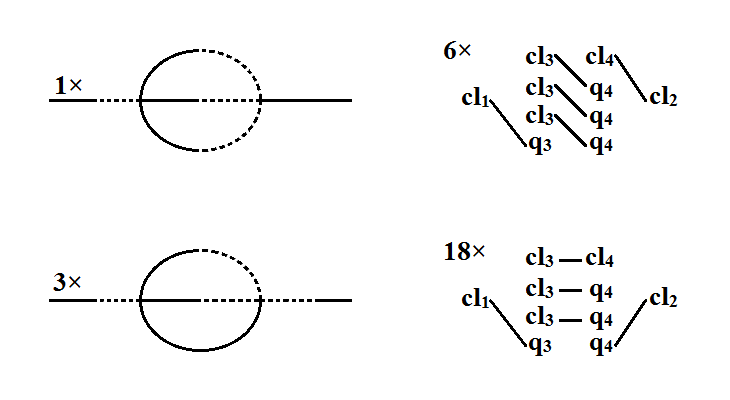}}\caption{`Sunset' diagram corrections to the Keldysh propagator (on the left side) and corresponding expressions (on the right side)}\label{fig:new-technique-2}
\end{figure}

To restore the correct combinatoric factors one can use more accurate diagrams corresponding to the correlation functions. These diagrams can be obtained from the standard diagrams by splitting all the vertices according to the indices of the outgoing lines and then contracting them in all possible ways again (Fig.~\ref{fig:new-technique-1}). Of course, these diagrams are nothing more than a convenient way to write down the expressions of the form~\eqref{eq:path-decomposition} and~\eqref{eq:contractions} and count the number of equal contractions. Note that this approach restores the relative ratios following from the symmetry of the standard diagrams, but also assigns to them an additional correct combinatoric factor. For example, this factor equals 3 in the case of Fig.~\ref{fig:new-technique-1} and equals 6 in the case of Fig.~\ref{fig:new-technique-2}.

Finally, let us stress that one has to take into account only connected diagrams in this technique. Indeed, all the parts of these diagrams which are not connected with external points inevitably contain the product of theta-functions of the form $\theta(t_1 - t_2) \theta(t_2 - t_3) \cdots \theta (t_n - t_1)$ which is equal to zero. One can prove it by induction: the simplest one-vertex diagram does contain such a product and expanding of the vertices (see Fig.~\ref{fig:vertex-1} and Fig.~\ref{fig:vertex-2}) only lengthens it. At the same time, diagrams where two disconnected parts are connected with different external points can not exist because there should be an even number of lines in both of the parts.

We will show that this technique gives the right result for one-loop and two-loop corrections in the following subsections~\ref{subsec:diagrams-one} and~\ref{subsec:diagrams-two}. For short we will draw and denote by the word `diagram' only standard diagrams with solid and dashed lines and restore the correct combinatoric factors in the final expressions.

\subsection{One-loop corrections}
\label{subsec:diagrams-one}

\begin{figure}[ht]
\center{\includegraphics[scale=0.75]{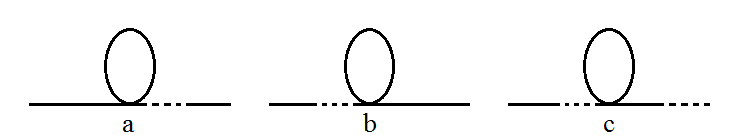}}\caption{All non-zero one-loop corrections to the Keldysh (a, b) and retarded (c) propagator}\label{fig:one-loop}
\end{figure}

Let us calculate one-loop corrections to the free theory propagators using this diagrammatic technique. In the Keldysh propagator we have six possible diagrams. They depend on times on the external legs $t_1$ and $t_2$ ($t_1 > t_2$), and the integration is performed over the intermediate time $t_3$. Nevertheless, we should calculate only two integrals (Fig. \ref{fig:one-loop}a and \ref{fig:one-loop}b) since all the others contain terms like $D_0^R(t_3, t_3)$ and obviously are equal to zero. Then, for the Keldysh propagator we have:
\begin{align}
D_1^K(t_1, t_2) &= D_{1,a}^K(t_1, t_2) + D_{1,b}^K(t_1, t_2) = \nonum
&= -3 i \lambda \int dt_3 \Big[ D_0^K(t_1, t_3) D_0^K(t_3, t_3) D_0^A(t_3, t_2) + D_0^R(t_1, t_3) D_0^K(t_3, t_3) D_0^K(t_3, t_2) \Big] = \nonum
&= -\frac{3 i \lambda}{2} \Big[ f_1 f_2^* \cdot \int_{T - \frac{\tau}{2}}^{T + \frac{\tau}{2}} dt_3 |f_3|^4 + f_1 f_2 \Big( 2 \int_{t_0}^{T} + \int_T^{T - \frac{\tau}{2}} + \int_T^{T + \frac{\tau}{2}} \Big) dt_3 |f_3|^2 f_3^{*2} - h.c. \Big] = \nonum
&= -9 i \lambda \cdot \frac{T}{(2 \omega_+)^2} \cdot (|\alpha|^2 + |\beta|^2) \Big( \alpha^* \beta^* f_1 f_2 - \alpha \beta f_1^* f_2^* \Big) + O(T^0).
\end{align}
We have the similar picture in cases of retarded and advanced propagators (Fig. \ref{fig:one-loop}c). Recall that $D_0^A(t_1, t_2) = D_0^R(t_2, t_1)$. Hence, we can consider only the retarded propagator:
\begin{align}
D_1^R(t_1, t_2) &= D_{1, c}^R(t_1, t_2) = -i \lambda \int dt_3 D_0^R(t_1, t_3) D_0^K(t_3, t_3) D_0^R(t_3, t_2) = \nonum
&= -i \lambda \int_{t_2}^{t_1} dt_3 \big( f_1 f_3^* - f_3 f_1^* \big) |f_3|^2 \big( f_3 f_2^* - f_2 f_3^* \big) = O(T^0).
\end{align}
These expressions are exactly the same as in the direct calculations performed in the previous section~\ref{sec:varfreq}.

\subsection{Two-loop corrections}
\label{subsec:diagrams-two}

Now let us calculate two-loop corrections. It is obvious that $D_2^R(t_1, t_2) \simeq O(T)$ and $D_2^A(t_1, t_2) \simeq O(T)$, because the corresponding integrals inevitably contain products of theta-functions of the form $\theta (t_4 - t_2) \theta (t_3 - t_4) \theta (t_1 - t_3)$ and hence they behave as $\simeq O(T\tau)$, i.e. are suppressed in the limit $T \rightarrow \infty$, $\lambda T = \text{const}$, $\tau = \text{const}$. This again coincides with the dependence obtained from the direct calculation as it was expected.

The second order correction to the Keldysh propagator is more sophisticated. Naively we have ten `sunset' diagrams, but diagrams $KKRAK$, $KRRAA$ and $RRAAK$ are obviously equal to zero (here we use such notation as in the Fig. \ref{fig:two-loop}). It is convenient to combine remaining diagrams with their time-reversals (e.g. $KAAAA$ and $RRRRK$) and integrate over intermediate times $t_3$ and $t_4$. Summing all the expressions and keeping only terms with non-oscillating integrands we obtain:
\begin{figure}[ht]
\center{\includegraphics[scale=0.75]{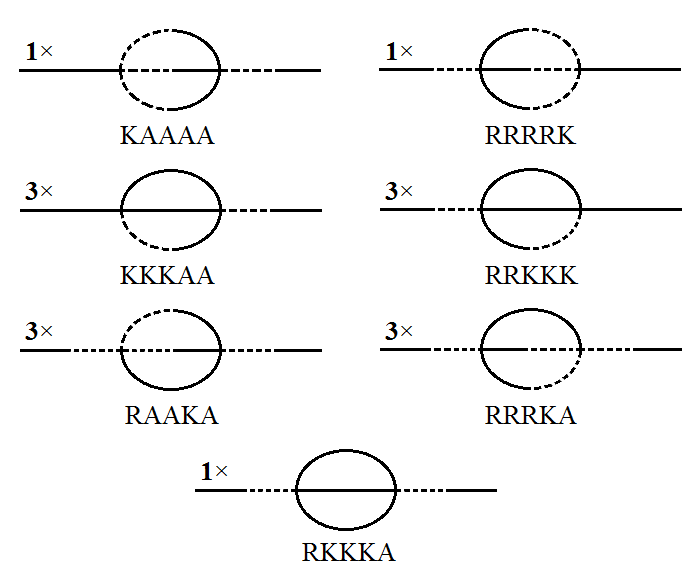}}\caption{All non-zero `sunset' two-loop corrections to the Keldysh propagator}\label{fig:two-loop}
\end{figure}
\begin{align}
\frac{1}{6 \lambda^2} D_2^K(t_1, t_2) &= + f_1 f_2^* \cdot \int_{t_0}^T d\eta f^4(\eta) \int_{t_0}^T d\xi f^{*4}(\xi) - \nonum &\phantom{=}- f_1 f_2 \cdot 2 \int_{t_0}^T d\eta f^2(\eta) |f(\eta)|^2 \int_\eta^T d\xi f^{*4}(\xi) + h.c. + O(T), \\
D_2^K(t_1, t_2) &= \frac{(\lambda T)^2}{(2 \omega_+)^4} \cdot 216 |\alpha|^2 |\beta|^2 \cdot \nonum &\phantom{=}\cdot \Big[ |\alpha|^2 |\beta|^2 \Big( f_1 f_2^* + f_1^* f_2 \Big) - \frac{1}{2} \big( |\alpha|^2 + |\beta|^2 \big) \Big( \alpha \beta f_1^* f_2^* + \alpha^* \beta^* f_1 f_2 \Big) \Big] + O(T). 
\end{align}
This result differs from the result obtained in direct calculations with the decomposition of the evolution operator~\eqref{eq:direct-propagators}. However, so far we have considered only `sunset' corrections whereas `tower' and `double tadpole' (Fig. \ref{fig:two-loop-extra}) corrections also give $\lambda^2$ dependence. Taking them into account\footnote{Careful calculations yield an additional coefficient of 6 for the `sunset' diagrams and 9 for the `tower' and `double tadpole' diagrams.} we obtain:
\begin{figure}[ht]
\center{\includegraphics[scale=0.6]{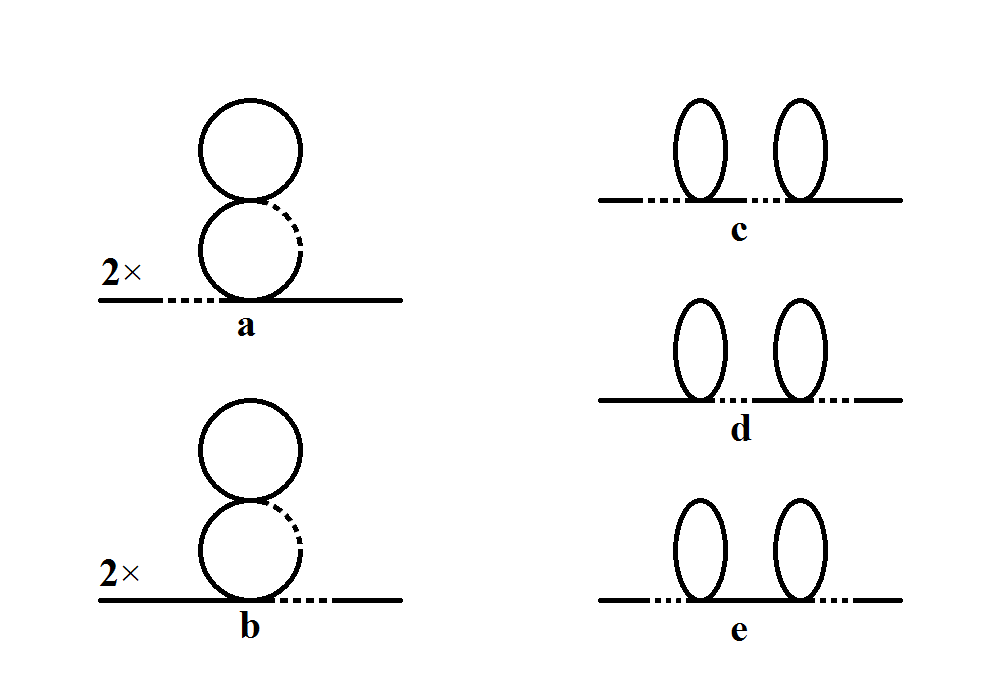}}\caption{Non-zero `tower' and `double tadpole' corrections to the Keldysh propagator}\label{fig:two-loop-extra}
\end{figure}
\begin{align}
\frac{1}{3\lambda^2} D_2^K(t_1, t_2) &= 
f_1 f_2^* \cdot \Bigg( 2 \int_{t_0}^T d\eta f^4(\eta) \int_{t_0}^T d\xi f^{*4}(\xi) + 3 \int_{t_0}^T d\eta f^2(\eta) |f(\eta)|^2 \int_{t_0}^T d\xi f^{*2}(\xi) |f(\xi)|^2 \Bigg) - \nonum
&- f_1 f_2 \cdot \int_{t_0}^T d\eta f^{*4}(\eta) \int_\eta^T d\xi f^2(\xi) |f(\xi)|^2 - \nonum
&- f_1 f_2 \cdot 9 \int_{t_0}^T d\eta f^{*2}(\eta) |f(\xi)|^2 \int_\eta^T d\xi |f(\xi)|^4 + h.c. + O(T).
\end{align}
And finally:
\begin{align}
D_2^K(t_1, t_2) &= \frac{(\lambda T)^2}{(2 \omega_+)^4} \cdot 27 \big( 3|\alpha|^4 + 3|\beta|^4 + 14 |\alpha|^2 |\beta|^2 \big) \cdot \nonum &\cdot \Big[ |\alpha|^2 |\beta|^2 \Big( f_1 f_2^* + f_1^* f_2 \Big) - \frac{1}{2} \big( |\alpha|^2 + |\beta|^2 \big) \Big( \alpha \beta f_1^* f_2^* + \alpha^* \beta^* f_1 f_2 \Big) \Big] + O(T).
\end{align}
This is the same expression as in the direct calculations~\eqref{eq:2nd}. `Tower' and `double tadpole' corrections to the R/A propagators are $\simeq O(T \tau)$ as well as `sunset' corrections. Thus one can see here that diagrammatic technique really works and gives exactly the same result as direct calculation for the ground state does.

Moreover, now one can see that the unnatural secular growth in the cases of constant and adiabatically varying frequency obtained in the subsections~\ref{subsec:constfreq} and~\ref{subsec:adfreq} is caused by `tower' and `double tadpole' diagrams. Indeed, taking into account that in this case $\beta = 0$ and recalculating corrections we obtain:
\begin{align}
D_2^{K, tadpoles}(t_1, t_2) &= \frac{27 i \lambda^2}{(2 \omega_-)^3} \Big( \frac{T}{(2 \omega_+)^2} - \frac{t_0}{(2 \omega_-)^2} \Big) \Big( e^{2 i \omega_- t_0} f_1^* f_2^* - e^{-2 i \omega_- t_0} f_1 f_2 \Big) + O(T^0), \nonum
D_2^{K, sunset}(t_1, t_2) &= D_2^{R/A, tadpoles}(t_1, t_2) = D_2^{R/A, sunset}(t_1, t_2) = O(T^0), \nonum
D_1^K(t_1, t_2) &= D_1^{R/A}(t_1, t_2) = O(T^0),
\end{align}
where we distinguish contributions from the diagrams of the form Fig.~\ref{fig:two-loop-extra} and Fig.~\ref{fig:two-loop}. In fact, this result can be generalized: growth of the corrections to the propagators in an arbitrary order corresponds to the `multiple tadpole' diagrams which give $D_n^K(t_1, t_2) \sim T^{n - 1} \cdot O(T^0)$, $D_n^{R/A}(t_1, t_2) = O(T^0\tau^n) = O(T^0)$.

\subsection{Corrections to the vertices}
\label{subsec:diagrams-vertex}

So far we have calculated corrections to the propagators. However, there can be the other type of corrections to the vertices. Here we will show that these corrections are suppressed in the limit $T \rightarrow \infty$, $\tau = \text{const}$. In other words, we set the times of the external legs $t_1, t_2, t_3, t_4$ to infinity while keeping their differences $t_1 - t_2$, etc. constant. We denote such differences as $\tau = t_i - t_j$.

\begin{figure}[ht]
\center{\includegraphics[scale=0.75]{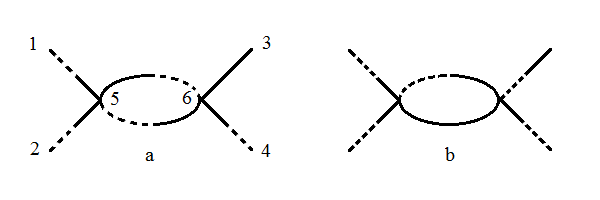}}\caption{One-loop corrections to the vertex $-i \lambda / 4$}\label{fig:vertex-1}
\end{figure}

First, let us consider one-loop ($\sim \lambda^2$) correction to the vertex with three dashed legs ($-i \lambda / 4$). It is easy to see that one of two possible diagrams in this case is equal to zero, because it contains the product of the theta-functions $\theta(t_5 - t_6) \theta(t_6 - t_5)$ (Fig. \ref{fig:vertex-1}). The other one contains $\theta(t_5 - t_1) \theta(t_6 - t_5) \theta(t_6 - t_4) \theta(t_3 - t_6)$ and hence it is $\sim \tau^2$, i.e. it is $O(T^0)$.

\begin{figure}[ht]
\center{\includegraphics[scale=0.65]{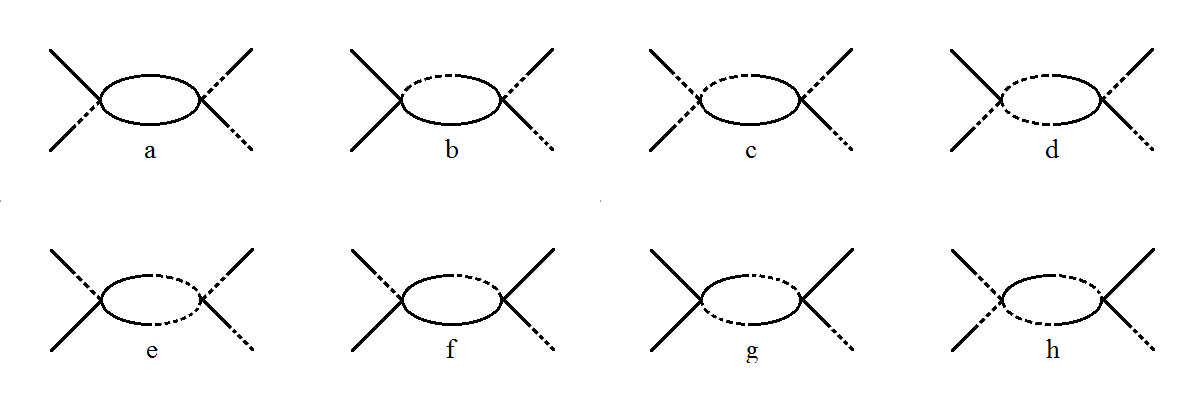}}\caption{One-loop corrections to the vertex $-i \lambda$}\label{fig:vertex-2}
\end{figure}

The correction to the other vertex is more complicated (Fig. \ref{fig:vertex-2}). Nevertheless, it is not difficult to estimate the contribution of each diagram separately.

Diagrams a -- d contain terms of the form $D^A(t_6, t_3) D^R(t_6, t_4) \sim \theta(t_3 - t_6) \theta(t_6 - t_4)$ (these terms correspond to the right legs) and hence they are $\sim \tau$. Though, they still have integrations of the form $\int_{t_0}^T$, which in the case $\beta \ne 0$ gives $O(T \tau)$ dependence. However, in the case $\beta = 0$ this integration can give only the expressions of the form $O(T^0)$ and the contributions of these diagrams are $\simeq O(T^0\tau)$. Diagrams e, f contain integration of $\theta(t_1 - t_5) \theta(t_5 - t_6) \theta(t_6 - t_4)$ and hence they are $\simeq O(T^0 \tau^2)$. Finally, diagrams g and h are obviously equal to zero.

In all we obtain that corrections to the vertices are $\sim \lambda^2 T$ in the non-adiabatically varying frequency case $\beta \ne 0$. These corrections grow with time but are suppressed in comparison with the corrections to the Keldysh propagators $D_1^K$ and $D_2^K$ in the limit $T \rightarrow \infty$, $\tau = \text{const}$. In the adiabatically varying frequency case $\beta = 0$ corrections to the vertices are $\sim \lambda^2 T^0$ and also are suppressed in comparison with corrections to the propagators in the limit under consideration.

\subsection{Renormalization of the frequency}
\label{subsec:renorm}

Above we have seen that in the adiabatic case $\beta = 0$ one can remove the secular growth of the corrections to the Keldysh propagator introducing the new vacuum state which tends to the vacuum state of the free Hamiltonian on the past and future infinities (see subsection~\ref{subsec:causes}). This is due to the fact that diagonal terms refer to the tadpole diagrams (see subsection~\ref{subsec:diagrams-two}) and behave as $\sim \lambda^n T^{n-1}$, when $T \rightarrow \infty$ whereas the off-diagonal terms behave as $\sim \lambda^n T^n$ in the same limit. However, in the non-adiabatic case $\beta \ne 0$ both diagonal and off-diagonal expressions grow as $\sim \lambda^n T^n$ and one cannot remove this growth by such a redefinition of the vacuum state.

In this subsection we show that `tadpole' one-loop corrections to the Keldysh propagator, which grow as $T \rightarrow \infty$, can be removed by a renormalization of the frequency~\cite{Akhmedov:2013-phi3, Akhmedov:2013-phi4, Peskin, Fetter, Wilson}. It is more convenient to prove this statement before the Keldysh rotation~\eqref{eq:keldysh-rotation}. In fact, let us sum all the contributions of the `tadpole' diagrams to the propagator $D^{++}$ using the Dyson-Schwinger equation~\cite{Landau:vol10}:
\beq
\label{eq:DS-renorm}
D_{exact}^{++}(t_1, t_2) = D_0^{++}(t_1, t_2) - 3 i \lambda \int dt_3 \, M^2(t_3) \Big[ D_0^{++}(t_1, t_3) D_{exact}^{++}(t_3, t_2) - D_0^{+-}(t_1, t_3) D_{exact}^{-+}(t_3, t_2) \Big],
\eeq
where we denoted as $M^2(t_3) \equiv D_0^{++}(t_3, t_3) = D_0^{--}(t_3, t_3) = |f_3|^2$. This summation corresponds to the so-called `tadpole chain' (Fig.~\ref{fig:Dyson}). Acting on the exact tadpole propagator $D_{exact}^{++}$ by the operator $(\partial_t^2 + \omega^2)$ and using the fact that the operator $\hat{D}_0$ is the inverse of~\eqref{eq:inversedG} we obtain:
\begin{figure}[ht]
\center{\includegraphics[scale=0.6]{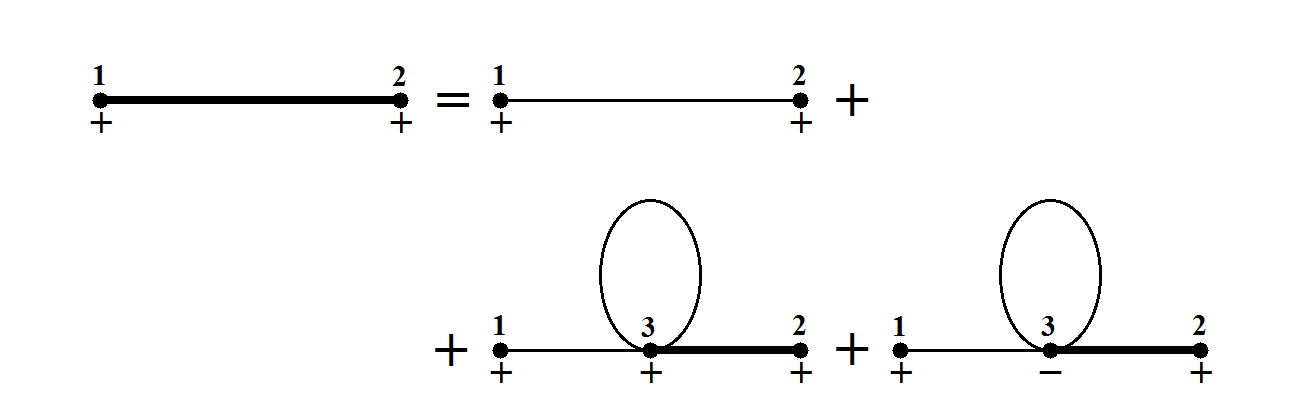}}\caption{`Tadpole chain' corrections to the $D^{++}$ propagator. Thin lines correspond to the bare propagators $D_0^{++}$ and $D_0^{+-}$, thick lines --- to the exact propagators $D^{++}$ and $D^{-+}$}\label{fig:Dyson}
\end{figure}
\beq \big[ \partial_{t_1}^2 + \omega^2(t_1) \big] D_{exact}^{++}(t_1, t_2) = \delta(t_1 - t_2) + 3 \lambda M^2(t_1) D_{exact}^{++}(t_1, t_2). \eeq
Introducing $\tilde{\omega}^2(t) = \omega^2(t) - 3 \lambda M^2(t)$ we obtain that:
\beq
\label{eq:DS-first}
\big[ \partial_t^2 + \tilde{\omega}^2(t) \big] D_{exact}^{++}(t, t') = \delta(t - t').
\eeq
Similar calculations for the other propagators give:
\begin{gather}
\big[ \partial_t^2 + \tilde{\omega}^2(t) \big] D_{exact}^{\pm \mp}(t, t') = 0, \nonum
\label{eq:DS-gather}
\big[ \partial_t^2 + \tilde{\omega}^2(t) \big] D_{exact}^{--}(t, t') = -\delta(t - t').
\end{gather}
Thus, `tadpole' corrections can be interpreted as the renormalization of the frequency in the action~\eqref{eq:varaction} or~\eqref{eq:varaction-path}.

In fact, let us consider the pure `tadpole' corrections to the Keldysh propagator in the several lowest orders in $\lambda$ in the case of non-adiabatically varying frequency $\beta \ne 0$. Substituting the behaviour of the modes at the future infinity~\eqref{eq:fcases} we obtain:
\begin{align}
D_0^K &= \frac{1}{2 \omega} \Big[ \alpha \beta^* e^{2 i \omega T} + \alpha^* \beta e^{-2 i \omega T} \Big] + \big( |\alpha|^2 + |\beta|^2 \big) \cdot \frac{1}{2 \omega} \cos(\omega \tau), \nonum
D_1^K &= -\frac{9 i \lambda T}{(2 \omega)^2} \big( |\alpha|^2 + |\beta|^2 \big) \cdot \frac{1}{2 \omega} \Big[ \alpha \beta^* e^{2 i \omega T} - \alpha^* \beta e^{-2 i \omega T} \Big] + O(T^0), \nonum
D_2^K &= \frac{1}{2} \Big( -\frac{9 i \lambda T}{(2 \omega)^2} \big( |\alpha|^2 + |\beta|^2 \big) \Big)^2 \cdot \frac{1}{2 \omega} \Big[ \alpha \beta^* e^{2 i \omega T} + \alpha^* \beta e^{-2 i \omega T} \Big] + O(T).
\end{align}
One can notice that the sign before the exponential function with negative frequency is positive in the even order corrections and negative in the odd order. So we can sum all the one-loop corrections in the leading order as $T \rightarrow \infty$:
\beq D_{exact}^K \simeq \frac{1}{2 \tilde{\omega}} \Big[ \alpha \beta^* e^{2 i \tilde{\omega} T} + \alpha^* \beta e^{-2 i \tilde{\omega} T} \Big] + \big( |\alpha|^2 + |\beta|^2 \big) \cdot \frac{1}{2 \tilde{\omega}} \cos(\tilde{\omega} \tau) + O(\lambda), \eeq
where we introduced the renormalized frequency
\beq
\label{eq:renorm-freq}
\tilde{\omega} = \omega - \frac{9 \lambda}{8 \omega^2} \big( |\alpha|^2 + |\beta|^2 \big), \quad \text{i.e.} \quad \tilde{\omega}^2 \simeq \omega^2 - \frac{9 \lambda}{4 \omega} \big( |\alpha|^2 + |\beta|^2 \big),
\eeq
and replaced $\omega$ with $\tilde{\omega}$ in the denominator to the accuracy of $O(\lambda)$. This additional expression is suppressed in the limit $T \rightarrow \infty$, $\lambda \rightarrow 0$, $\lambda T = \text{const}$.

Note that the renormalized frequency obtained from the Dyson-Schwinger equation~\eqref{eq:DS-renorm}--\eqref{eq:DS-gather} does not coincide with the result of direct calculations~\eqref{eq:renorm-freq}. This is due to the fact that eq.~\eqref{eq:DS-renorm} corresponds only to the `tadpole chain' diagrams of the form Fig.~\ref{fig:one-loop}c, Fig.~\ref{fig:two-loop-extra}c--e and so on, instead of the general `multiple tadpole' diagrams including `tower' (Fig.~\ref{fig:two-loop-extra}a,b) and more complex contributions.

In the case of adiabatically varying frequency $\beta = 0$ one-loop corrections grow slower:
\begin{align}
\label{eq:renorm-ad}
D_0^K &= \frac{1}{2 \omega_+} \cos(\omega_+ \tau), \nonum
D_1^K &= \frac{3 \lambda}{(2 \omega_-)^2} \cdot \frac{1}{4 \omega_+ \omega_-} \Big[ e^{2 i \omega_+ T - 2 i \omega_- t_0} + e^{-2 i \omega_+ T + 2 i \omega_- t_0} \Big], \nonum
D_2^K &= \frac{3 \lambda}{(2 \omega_-)^2} \cdot (-9 i \lambda) \Big( \frac{T}{(2 \omega_+)^2} - \frac{t_0}{(2 \omega_-)^2} \Big) \cdot \frac{1}{4 \omega_+ \omega_-} \Big[ e^{2 i \omega_+ T - 2 i \omega_- t_0} - e^{-2 i \omega_+ T + 2 i \omega_- t_0} \Big] + O(T^0).
\end{align}
One can see that the bare propagator $D_0^K$ depends only on the difference $\tau$ whereas the corrections $D_1^K$, $D_2^K$ and so on also depend on the average time $T$. Hence, one cannot obtain the expression of the form $D_0^K$ after the summation of all such corrections as~\eqref{eq:renorm-ad}. This fact corresponds to the breakdown of the WKB approximation after the renormalization of the frequency. Indeed, let us consider the lowest order corrections to the mass term in~\eqref{eq:DS-renorm}:
\beq M^2(t) = D^K(t, t) = \frac{1}{2 \omega_+} + \frac{3 \lambda}{8 \omega_- \omega_+^3} \cos(2 \omega_+ t) + \frac{27 \lambda^2}{8 \omega_- \omega_+^3} \Big( \frac{t}{(2 \omega_+)^2} - \frac{t_0}{(2 \omega_-)^2} \Big) \sin(2 \omega_+ t) + \cdots, \eeq 
and check the WKB approximation~\eqref{eq:WKB} for the frequency from the eq.~\eqref{eq:DS-first}:
\beq \Big| \frac{d}{dt} \frac{1}{\tilde{\omega}} \Big| = \Big| \frac{\dot{\tilde{\omega}}}{\tilde{\omega}^2} \Big| = \Big| \frac{\dot{\tilde{\omega}}}{\omega^2(t) - 3 \lambda M^2(t)} \Big| \ll 1. \eeq
One can see that $M^2(t) \rightarrow \infty$ as $t \rightarrow \infty$ while $\omega^2(t)$ remains finite in such a limit, so at least one turning point $\omega^2(t_*) = 3 \lambda M^2(t_*)$ appears when $\lambda \ne 0$. Hence, the inequality cannot be satisfied for all times $t \in (-\infty, +\infty)$ and the WKB approximation breaks down. Thus, one cannot hide the growth of `tadpole' corrections into the renormalization of the frequency in this case. However, we remind that in the subsection~\ref{subsec:causes} we have shown that this correction can be removed by a modification of the vacuum state (in that case $\omega = \text{const}$, so we set $\omega_+ = \omega_-$ and $t_0 = 0$). Also note that in the non-adiabatically varying frequency case one does not have such problems because inequality~\eqref{eq:WKB} does not hold even when $\lambda = 0$, and therefore the renormalization of the frequency does not change the behaviour of the propagators. 

Thus, in this subsection we have shown that one can hide the growth of `tadpole' corrections to the Keldysh propagator into a renormalization of the frequency in the non-adiabatically varying frequency case, but cannot do the same in the adiabatically varying frequency case. Note, however, that higher than second order corrections in the non-adiabatically case still grow because they correspond to the `sunset' diagrams. At the same time, such a growth corresponds to the time dependence of $\langle a^+ a \rangle$ and $\langle a a \rangle$ which are components of the Keldysh propagator~\eqref{eq:nk}.

\section{Discussion and acknowledgements}
\label{sec:conclusion}

In this paper we have considered the simplest example of a non-stationary quantum field theory which is quantum mechanical oscillator with varying frequency and $\lambda \phi^4$ self-interaction. We used non-equilibrium diagrammatic technique to calculate the loop corrections to the propagators and vertices in the ground state of the free Hamiltonian and checked the results by direct calculations in Hamiltonian quantization in the interaction picture.

First, we obtained that leading loop corrections in the limit $T \rightarrow \infty$ in adiabatically varying frequency case show the same behaviour as in the stationary, i.e. constant frequency case. In both these cases corrections to the Keldysh propagator grow indefinitely with time whereas corrections to the retarded/advanced propagators and vertices remain finite. This growth refers only to the `tadpole' diagrams. However, one can remove the growth by a modification of the vacuum state which tends to the identical transformation on the future and past infinities. Hence, neither level population $n = \langle a^+ a \rangle$ nor anomalous quantum average $\kappa = \langle a a \rangle$, which are components of the Keldysh propagator, grow in this case. This fact is related to the well-known adiabatic theorem in quantum mechanics which was first formulated in 1928 by Max Born and Vladimir Fock~\cite{Born}.

Second, in the non-adiabatically varying frequency case loop corrections to the Keldysh propagator grow faster than in the adiabatic case and cannot be removed by a modification of the vacuum state. Corrections to the retarded/advanced propagators and vertices also grow indefinitely but are suppressed in comparison with the Keldysh propagator by the higher powers of $\lambda$ in the limit $T \rightarrow \infty$, $\lambda \rightarrow 0$, $\lambda T = \text{const}$. This growth refers to both the `tadpole' and `sunset' diagrams. Despite the fact that one can hide the growth of the `tadpole' diagrams into a renormalization of the frequency, one cannot do the same with `sunset' diagrams. This means that both level population $\langle a^+ a \rangle$ and anomalous quantum average $\langle a a \rangle$ grow indefinetely in this case. This fact is strongly related to the non-adiabaticity of the frequency which brings energy into the system.

Finally, the secular growth just means the breakdown of the perturbation theory, because even if $\lambda$ is small $\lambda T$ can become of the order of unity. This means that one has to resum the leading corrections from all the loops and write down the kinetic equation to understand the behaviour of $n$ and $\kappa$. In quantum field theory one usually derives this equation using a diagrammatic approach and summing loop corrections in an arbitrary state of the theory~\cite{Kamenev, Berges, Rammer, Akhmedov:2013-phi3, Akhmedov:2013-phi4, Polyakov:2012, Akhmedov:2012}. Unfortunately, in quantum mechanics this approach works only for the vacuum (and thermal) state due to the absence of the infinite spatial volume. This means that one has to derive an analog of the kinetic equation in a different way, which will be done elsewhere. 

The authors would like to thank Andrey Semenov, Daniil Sherstnev and Anna Radovskaya for useful comments and discussions. Also we thank Emil T. Akhmedov for formulating the problem and sharing of his ideas. This work was done under the financial support of the Russian state grant Goszadanie 3.9904.2017/8.9.

\end{document}